\documentclass[twocolumn]{aastex631} %%for arxiv

\usepackage{amsmath}
\usepackage{xcolor, soul}
\usepackage{ulem}
\newcommand\redout{\bgroup\markoverwith{\textcolor{red}{\rule[0.5ex]{2pt}{0.8pt}}}\ULon}

\hypersetup{linkcolor=cyan,citecolor=teal,filecolor=cyan,urlcolor=blue}

\shorttitle{$L_{\rm MIR}$-based-$L_{\rm bol}$ and -$M_{\rm BH}$ estimators}
\shortauthors{Kim et al.}

\begin{document}

\title{Estimators of Bolometric Luminosity and Black Hole Mass\\
	with Mid-infrared Continuum Luminosities for Dust-obscured Quasars:\\
	Prevalence of Dust-obscured SDSS Quasars
	\footnote{dh.dr2kim@gmail.com (DK); myungshin.im@gmail.com (MI)}}

\author[0000-0002-6925-4821]{Dohyeong Kim}
\affiliation{Department of Earth Sciences, Pusan National University, Busan 46241, Republic of Korea}

\author[0000-0002-8537-6714]{Myungshin Im}
\affiliation{Astronomy Program, Dept. of Physics \& Astronomy, Seoul National University, Seoul 08826, Republic of Korea}
\affiliation{SNU Astronomy Research Center (SNU ARC), Astronomy Program, Dept. of Physics \& Astronomy, Seoul National University, Seoul 08826, Republic of Korea}

\author[0000-0002-3560-0781]{Minjin Kim}
\affiliation{Department of Astronomy and Atmospheric Sciences, College of Natural Sciences, Kyungpook National University, Daegu 41566, Republic of Korea}

\author[0000-0003-1647-3286]{Yongjung Kim}
\affiliation{Department of Astronomy and Atmospheric Sciences, College of Natural Sciences, Kyungpook National University, Daegu 41566, Republic of Korea}

\author[0000-0002-2188-4832]{Suhyun Shin}
\affiliation{Astronomy Program, Dept. of Physics \& Astronomy, Seoul National University, Seoul 08826, Republic of Korea}
\affiliation{SNU Astronomy Research Center (SNU ARC), Astronomy Program, Dept. of Physics \& Astronomy, Seoul National University, Seoul 08826, Republic of Korea}

\author[0000-0002-4179-2628]{Hyunjin Shim}
\affiliation{Department of Earth Science Education, Kyungpook National University, 80 Daehak-ro, Buk-gu, Daegu 41566, Republic of Korea}

\author[0000-0002-4362-4070]{Hyunmi Song}
\affiliation{Department of Astronomy and Space Science, Chungnam National University, 99 Daehak-ro, Yuseong-gu, Daejeon 34134, Republic of Korea}

\begin{abstract}

We present bolometric luminosity ($L_{\rm bol}$) and black hole (BH) mass ($M_{\rm BH}$) estimators
based on mid-infrared (MIR) continuum luminosity (hereafter, $L_{\rm MIR}$) that are measured from infrared (IR) photometric data.
The $L_{\rm MIR}$-based estimators are relatively immune from dust extinction effects,
hence they can be used for dust-obscured quasars.
To derive the $L_{\rm bol}$ and $M_{\rm BH}$ estimators,
we use unobscured quasars selected from the Sloan Digital Sky Survey (SDSS) quasar catalog,
which have wide ranges of $L_{\rm bol}$ ($10^{44.62}$--$10^{46.16}$\,$\rm erg\,s^{-1}$)
and $M_{\rm BH}$ ($10^{7.14}$--$10^{9.69}$\,$M_{\odot}$).
We find empirical relations between
(i) continuum luminosity at 5100\,$\rm{\AA{}}$ (hereafter, L5100) and $L_{\rm MIR}$;
(ii) $L_{\rm bol}$ and $L_{\rm MIR}$.
Using these relations, we derive the $L_{\rm MIR}$-based $L_{\rm bol}$ and $M_{\rm BH}$ estimators.
We find that our estimators allow the determination of $L_{\rm bol}$ and $M_{\rm BH}$
at an accuracy of $\sim$0.2\,dex against the fiducial estimates based on the optical properties
of the unobscured quasars.
We apply the $L_{\rm MIR}$-based estimators to SDSS quasars at $z \lesssim 0.5$ including obscured ones.
The ratios of $L_{\rm bol}$ from the $L_{\rm MIR}$-based estimators to those from the optical luminosity-based estimators 
become larger with the amount of the dust extinction,
and a non-negligible fraction ($\sim$15\,\%) of the SDSS quasars exhibits ratios greater than 1.5.
This result suggests that dust extinction can significantly affect physical parameter derivations even for SDSS quasars,
and that dust extinction needs to be carefully taken into account when deriving quasar properties.

\end{abstract}


\section{Introduction} \label{sec:intro}

 In recent years, supermassive black holes (SMBHs)
have been found at the centers of spheroidal galaxies. 
In many observational studies,
the masses of SMBHs are well known to have correlations with the properties of their host galaxies, 
e.g., luminosities \citep{graham07a,bentz09,gultekin09,bennert10,greene10},
stellar velocity dispersions ($\sigma_{\rm \ast}$;
\citealt{ferrarese00,gebhardt00,tremaine02,gultekin09,woo10}),
and S\'ersic or concentration indices \citep{graham01,graham07b} of their spheroids.
These empirical relationships imply
the coevolution of SMBHs and galaxies (for review, see \citealt{kormendy13}).

Although the detailed physical mechanisms of the coevolution are still unknown,
SMBHs have been considered to play an important role in galaxy growth.
These SMBHs are believed to grow by accreting gas \citep{lynden-bell69} in an active phase,
i.e., the period of active galactic nuclei (AGNs),
and quasars refer to the ultraluminous active phase.

Although quasars emit their enormous energy in all wavelengths (i.e., gammaray to radio),
X-ray, ultraviolet (UV), optical, and radio observations have been mainly used to discover quasars.
Based on these surveys, 
previous studies \citep{grazian00,becker01,anderson03,croom04,risaliti05,schneider05,veron-cetty06,paris14,lyke20}
have found $\sim$1 million quasars to date.

However, there is a possibility that the previous quasar surveys
can miss a large number ($<$50\,\%)
of red-colored quasars \citep{comastri01,tozzi06,polletta08}.
Here, red quasars refer to quasars with red continua from the optical to near-IR
(NIR; e.g., $r^{\prime} - K > 5$\,mag; \citealt{urrutia09}), 
and red quasars can include both type 1, type 2, and extremely red quasars (e.g., \citealt{hamann17}). 
Many hot dust-obscured galaxies (\citealt{eisenhardt12,wu12}) also fall into this category.
Their red colors are suspected to come from
the extinction effects of the intervening material in their host galaxies \citep{webster95,cutri01}.
These red quasars have been considered a different population from normal quasars (i.e., unobscured quasars).

 Several simulation studies (e.g., \citealt{menci04,hopkins05,hopkins06,hopkins08})
predicted red quasars can be an intermediate population between gas-rich merger-driven star-forming galaxies,
often seen as ultraluminous infrared galaxies (\citealt{sanders88,sanders96}), 
and unobscured quasars after sweeping away the dust and gas by the quasar-driven outflow.
From this point of view, red quasars are young and dust-obscured quasars,
in which dust obscuration is due to the remaining gas and dust in their host galaxies.

 Several pieces of observational evidence have supported this scenario.
For example, red quasars have
(i) dusty red colors \citep{kim18a},
(ii) high Eddington ratios ($\lambda_{\rm Edd} = L_{\rm bol}/L_{\rm Edd}$,
where $L_{\rm bol}$ and $L_{\rm Edd}$ are the bolometric and Eddington luminosities, respectively; 
\citealt{urrutia12,kim15a,kim18a,kim18b,kim22}),
(iii) high fractions of merging features \citep{urrutia08,glikman15},
(iv) prospective merging SMBH candidates \citep{kim20},
and {\tiny }(v) enhanced star-formation activities \citep{georgakakis09},
as expected in the merger-driven galaxy evolution scenario.

 However, the interpretation of red quasars is still controversial
since several studies have provided different explanations for red quasars.
For example, \cite{wilkes02} and \cite{rose13} explained that the red colors of red quasars come from
a moderate viewing angle of dust torus in the unification model \citep{antonucci93}.
Moreover, other studies \citep{puchnarewicz98,whiting01,maddox06,young08,rose13,ruiz14}
suggested that red quasars have intrinsically red continua without any dust extinction. 
Also, an unusual hot dust covering factor \citep{rose14}
and a synchrotron emission peak at NIR wavelength \citep{whiting01}
were proposed as alternative explanations for the red colors of red quasars.

 To investigate whether red quasars are dust-obscured quasars
as expected in the merger-driven galaxy evolution scenario,
$\lambda_{\rm Edd}$ is a key property.
However, the $\lambda_{\rm Edd}$ values of only $<100$ red quasars have been measured
(e.g., \citealt{urrutia12,kim15a,kim18a,kim18b,kim22}),
and the limited sample size can cause the controversy in the interpretation of red quasars.

 To obtain $\lambda_{\rm Edd}$, two fundamental physical quantities,
bolometric luminosities and BH masses, should be measured.
The bolometric luminosity is the total radiative energy in all wavelengths.
Measuring the bolometric luminosity accurately is an arduous task
since it needs full wavelength observation
at least from 1\,$\rm{\mu m}$ to 10\,keV or 100\,keV (e.g., \citealt{krawczyk13}).
However, the bolometric luminosities have empirical relations with various monochromatic continuum luminosities
at UV and optical (e.g., 1450\,$\rm \AA{}$ and 5100\,$\rm \AA{}$;
\citealt{elvis94,kaspi00,richards06,runnoe12}),
and these relationships have been widely used.

 The BH masses can be measured by using
velocity widths of broad lines and sizes of broad-line regions (BLRs).
The velocity width can be parameterized by full width at half maximum (FWHM).
The size of BLR can be determined from a reverberation mapping technique \citep{peterson04},
but the reverberation mapping method is time-consuming.
Therefore, the BLR size has been mainly estimated from a single-epoch spectrum
based on empirical relations between the BLR sizes and UV or optical continuum luminosities (e.g., \citealt{kaspi05}).

 However, despite that the FWHM values even measured in UV and optical cannot be affected by the dust extinction effects,
the UV and optical continuum fluxes are easily affected by the extinction in dust-obscured systems.
Of course there is the rare possibility that the FWHM could change if the BLR is partially obscured by dust, but
\cite{kim18b} showed that the FWHM values of H$\beta$, H$\alpha$, P$\beta$, and P$\alpha$ lines of red quasars are not sufficiently different.
On the contrary, their Balmer line luminosities are significantly suppressed compared to their Paschen line luminosities. 
For example, if a red quasar is obscured by a color excess of $E(B-V)=1$\,mag \citep{glikman07,kim15a,kim18b}, 
its continuum fluxes at 5100\,$\rm \AA{}$ and 1450\,$\rm \AA{}$ are suppressed
by factors of 22.5 and 2100, respectively,
which is calculated by performing \texttt{FM$\_$UNRED} code \citep{fitzpatrick99}
with an assumption of $\it R_V=3.1$ (e.g., \citealt{weingartner01}).

 For red quasars, therefore, the two fundamental quantities ($L_{\rm bol}$ and $M_{\rm BH}$)
estimated from the UV and optical indicators can be easily affected by the dust extinction.
To alleviate the dust extinction effects, 
the Paschen- and Brackett-line-based $M_{\rm BH}$ \citep{kim10,kim15a,kim15b}
and $L_{\rm bol}$ \citep{kim22} estimators were derived.
However, the Paschen and Brackett line properties can be achieved through IR spectroscopic observations
that have been rarely performed.

In contrast, IR photometry has been extensively obtained by several large area surveys,
such as the Two Micron All Sky Survey (2MASS; \citealt{cutri03}) and Wide-field Infrared Explorer ($\it WISE$; \citealt{wright10}).
The IR photometry represents the IR continuum shape dominated by dust component radiations (e.g., \citealt{kim15b}).
The AGN dust emission arises from a dust torus 
that is believed to surround the accretion disk and SMBH in the Unification model \citep{antonucci93}.
The dust component can be broadly divided into two subcomponents depending on their temperatures,
and the two subcomponents are hot ($>1000$\,K; e.g., \citealt{barvainis87}) and warm ($\sim200$\,K; e.g., \citealt{netzer07}) dust components.
The hot dust component radiation peaks at $\sim$2--3\,$\mu$m,
which is known to dominate mid-IR (MIR) quasar continuum (e.g., \citealt{kim15b}).

 MIR continuum luminosities have been known to be used as
the $L_{\rm bol}$ and $M_{\rm BH}$ estimators.
This approach is supported by several observational studies that
(i) the hot dust luminosities and the dust torus sizes have empirical correlations with the bolometric luminosities \citep{kim15b,suganuma06};
(ii) the bolometric luminosities are correlated with the BLR sizes \citep{kaspi00}.

 The $L_{\rm MIR}$-based estimators can be used for dust-obscured systems.
The MIR continuum luminosities are relatively immune from the dust extinction.
For example, for the case of a quasar at $z = 0$ of which $E(B-V) =1$\,mag, %in the case of $E(B-V)=1$\,mag, 
its fluxes at the $\it W1$- (3.4\,$\mu$m) and $\it W2$- (4.6$\mu$m) bands
only decrease by factors of 1.20 and 1.14, respectively.
Note that this effect is dependent on redshift, but the $\it{W1}$- and $\it{W2}$-band fluxes of low-$z$ ($\lesssim 0.5$) quasars
are still relatively immune from dust extinction.
If we assume a quasar at $z=0.5$ and with $E(B-V)=1$, the $\it{W1}$- and $\it{W2}$-band wavelengths in the rest-frame are 2.27 and 3.07\,$\mu$m, respectively,
and their continuum fluxes also only decrease by factors of 1.35 and 1.23, respectively.
 
 Furthermore, the MIR continuum luminosities can be used as common indicators
for both unobscured and red quasars.
\cite{kim18a} showed that ratios of the hot dust luminosities to the bolometric luminosities
for the two kinds of quasars are not significantly different.
Although \cite{ichikawa19} reported that 
the hot dust luminosity ratios of the two types of quasars can be different,
there have been several results that the different ratios can be introduced from a selection bias \citep{elitzur12,lanz19}.

 In this paper, we find empirical relationships between the bolometric luminosities
and MIR continuum luminosities. 
Moreover, we derive BH mass estimators
based on a combination of the optical-line FWHM values and the MIR continuum luminosities.
Since (i) the MIR continuum luminosities can be measured from the all-sky IR photometric data (2MASS and $\it WISE$) and
(ii) optical spectra are more accessible than IR spectra,
these $L_{\rm MIR}$-based estimators can be extensively used to investigate the nature of red quasars.
The $L_{\rm MIR}$-based $M_{\rm BH}$ estimators are derived using the FWHM of the broad-line component of H$\beta$ line,
and therefore, it can only be applied to type 1 red quasars.
Hereafter, red quasars refer to type 1 red quasars.
Moreover, we examine the importance of the dust extinction correction by applying the $L_{\rm MIR}$-based estimators 
to a general population of Sloan Digital Sky Survey (SDSS) quasars. 
We expect these estimators can also be applied to 
different types of dust-obscured quasars (e.g., intermediate-type and Compton thick quasars),
but the $M_{\rm BH}$ estimators would only be feasible when their broad-line component is visible.

Throughout this work, we use a standard $\Lambda$CDM model of
$H_{0}=70\,{\rm km\,s^{-1}}$\,Mpc$^{-1}$, $\Omega_{m}=0.3$, and $\Omega_{\Lambda}=0.7$.
This model has been supported by observational studies conducted
in recent years (e.g. \citealt{im97,planck16}).

\section{The Sample and Data} \label{sec:sample}
 Our sample is drawn from 
the SDSS Data Release 14 (DR14) quasar catalog \citep{paris18}.
The SDSS DR14 quasar catalog contains 526,265 spectroscopically confirmed quasars,
and they have a wide redshift range of 0.004--6.97.
Moreover, this catalog provides several IR photometric survey data, such as
the 2MASS Point Source Catalog (PSC; \citealt{cutri03}) and 
$\it{WISE}$ \citep{wright10,mainzer11} All-WISE Point Source Catalog \citep{cutri21}.

 \cite{rakshit20} measured the spectral properties of the SDSS DR14 quasars
by performing multicomponent fitting with \texttt{PyQSOFit} code \citep{guo19,shen19}.
They provide the measured FWHM values of several broad emission lines,
such as H$\alpha$, H$\beta$, \ion{Mg}{2}, \ion{C}{4}, and Ly$\alpha$.
Moreover, the BH masses are also measured
based on various FWHM values and continuum luminosities (e.g., \citealt{vestergaard06,shen11}),
expressed as
\begin{equation}
	\begin{aligned}
	\log & \left( \frac{M_{\rm BH}}{M_{\odot}} \right) =\\
	&\rm{\alpha}+\rm{\beta} \log \left( \frac{\lambda \it{L}_{\lambda}}{\rm{10^{44}\,erg~s^{-1}}} \right)
	+2 \log \left( \frac{\rm{FWHM}}{1000\,{\rm km~s^{-1}}} \right).\label{eqn:MBH_gen}
	\end{aligned}
\end{equation}

 We adopt the optical-line-based FWHM values and BH masses presented in \cite{rakshit20}.
The ${\rm FWHM_{H\beta}}$ and ${\rm FWHM_{H\alpha}}$
are adopted as the optical broad-line FWHM values.
Moreover, the BH masses were measured with
the $\rm{FWHM_{H\beta}}$ and L5100
by using 0.91 and 0.50 for the $\alpha$ and $\beta$ values in Equation~\ref{eqn:MBH_gen}, respectively \citep{vestergaard06}.

 Among the SDSS DR14 quasars, we select the quasars that satisfy the following condition: 
(1) being detected with good signal-to-noise ratio (S/N), (2) being unobscured, and 
(3) being brighter than a uniform detection limit.
We select quasars that are detected with an S/N$> 3$ in the $\it J$, $\it H$, $\it K$ bands (from 2MASS PSC), 
as well as in $\it W1$ and $\it W2$ bands (from $\it WISE$).

 SDSS quasars are known to also include dust-obscured quasars (e.g., \citealt{richards03}),
hence we choose unobscured quasars among the SDSS quasars.
First, the unobscured quasars are chosen by using broad line Balmer decrement.
The $ {L_{\rm H\alpha}} / {L_{\rm H\beta}} $ values of unobscured quasars
have been found to vary from 3.06 \citep{dong08}, 
which is similar to the value of 3.1 predicted by Case B recombination.
However, the Balmer decrements are distinct for radio-loud and -quiet quasars.
For instance, the $\log \left( \frac{L_{\rm H\alpha}}{L_{\rm H\beta}} \right)$ values were found to be
0.528$\pm$0.057 and 0.483$\pm$0.046 for radio-loud and -quiet quasars, respectively \citep{dong08},
and hence we consider the bimodality of the Balmer decrement to select unobscured quasars.
Second, we select the unobscured quasars via spectral energy distribution (SED) fitting.
We perform the SED fitting for the SDSS quasars as described in Section~\ref{sec:Lcont}.
The SED fitting yields a color excess, $E(B-V)$, and red quasars have been classified as
having high $E(B-V)$ values (e.g., $E(B-V)>0.1$; \citealt{urrutia09}).
In this work, we use the SDSS quasars with Balmer decrement and color excess ranges of 
(i) $(0.528 - 3 \times 0.057) < \log \left( \frac{L_{\rm H\alpha}}{L_{\rm H\beta}} \right) < (0.483 + 3 \times 0.046)$;
and (ii) $E(B-V) < 0.1$ as unobscured quasars.
Note that the Balmer decrement criterion can include some mildly obscured radio-loud quasars.
However, we expect that the remaining mildly obscured quasars are ultimately rejected by the $E(B-V)<0.1$ criterion.

 Moreover, we choose the quasars using a uniform detection limit
to eliminate the effects from the different detection limits of the SDSS, 2MASS, and $\it{WISE}$ data.
Since the 2MASS data have shallower detection limits (e.g., 14.3\,mag at the $K_{\rm s}$ band) than SDSS and $\it{WISE}$,
we measure the SDSS and $\it{WISE}$ magnitude limits that correspond to the detection limit of 2MASS.
In order to measure the magnitude limits, we assume a quasar at $z=0.25$ with $K_{\rm s}$-band magnitude of 14.3,
and measure its $\it{r}$- and $\it{W2}$-band magnitudes, where we use the AGN template of \cite{richards06}. 
The measured $\it{r}$- and $\it{W2}$-band magnitude limits are 17.14 (AB) mag and 11.48 (Vega) mag, respectively.
We select the SDSS quasars that are brighter than the magnitude limits of the $\it{r}$, $K_{\rm s}$, and $\it{W2}$ bands.

 We then match these quasars to the quasars in \cite{krawczyk13}, and adopt the bolometric luminosities.
For measuring the bolometric luminosities, the photons from $\sim$1\,$\mu$m, excluding IR emissions,
to $\sim$10\,keV have been widely used (e.g., \citealt{runnoe12,kim22}).
Hence, we adopt the 1\,$\mu$m--10\,keV luminosities measured in \cite{krawczyk13} as the bolometric luminosities.
Note that several previous studies (e.g., \citealt{richards06}) used 
different wavelength ranges (e.g., 100\,$\mu$m--10\,keV) for measuring the bolometric luminosities,
resulting in discrepancies in the bolometric luminosity measurements, % depending on the used wavelength range.
and the discrepancy is shown in Figure~\ref{fig:Lbol-LMIR}.

 \cite{krawczyk13} measured the 1\,$\mu$m--10\,keV luminosities of SDSS DR7 quasars using broadband data from the FUV to the MIR, 
and compensated for the lack of the X-ray data by applying an $L_{\rm UV}$--$L_{\rm X}$ relation \citep{steffen06}.
These bolometric luminosity measurements are reliable,
and the rms difference in the 1\,$\mu$m--10\,keV luminosities from \cite{krawczyk13} and \cite{richards06}
for 250 overlapped quasars was only 0.07\,dex \citep{kim22}.

\begin{figure}
	\centering
	\figurenum{1}
	\includegraphics[width=\columnwidth]{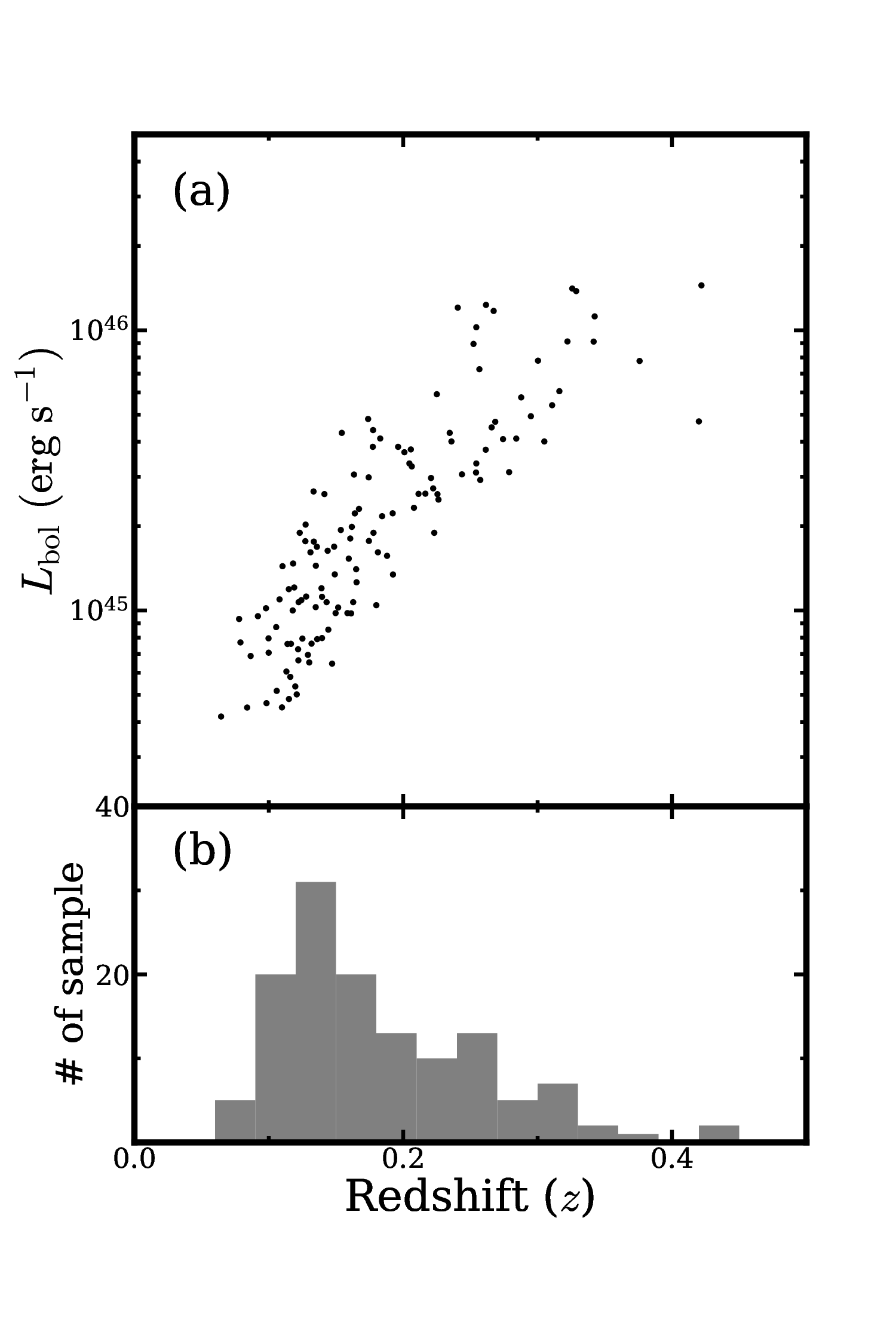}\\
	\caption{
		(a): Redshift versus bolometric luminosity. 
		The bolometric luminosities are adopted from \cite{krawczyk13}.
		(b): Redshift distribution of our sample.
		\label{fig:Sample1}}
\end{figure}

\begin{figure}
	\centering
	\figurenum{2}
	\includegraphics[width=\columnwidth]{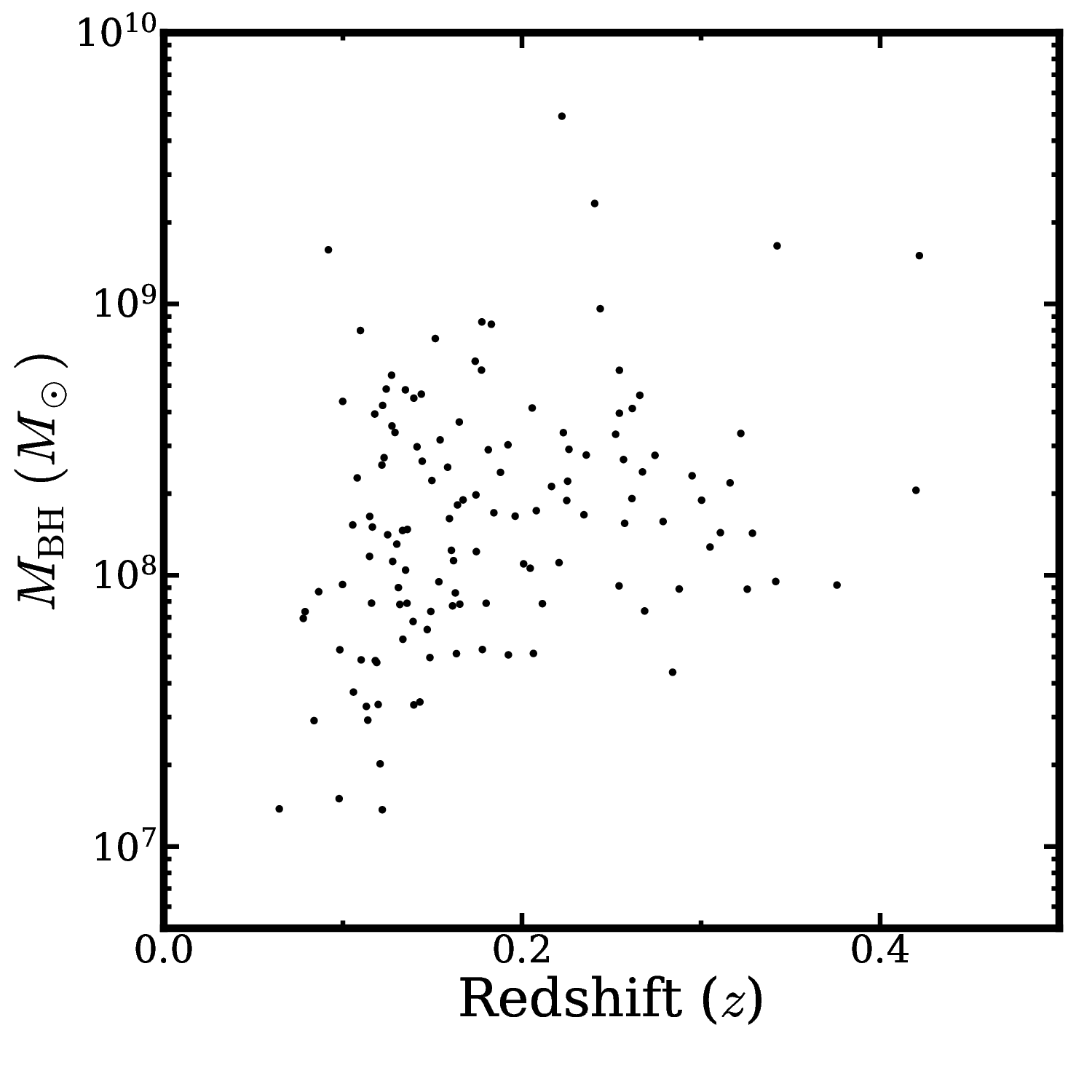}\\
	\caption{
		Redshift vs. BH mass for our sample.
		The BH masses are measured based on the L5100 and $\rm{FWHM_{H\beta}}$ values \citep{vestergaard06}.
		\label{fig:Sample2}}
\end{figure}

 Figures~\ref{fig:Sample1} and \ref{fig:Sample2} show
the basic properties of the selected 129 SDSS quasars.
Our sample is at low redshift ($0.06 <z< 0.42$), but
spans over somewhat wide ranges of BH mass 
($10^{7.14}\,M_{\odot}  < M_{\rm BH} < 10^{9.69}\,M_{\odot}$)
and bolometric luminosity 
($10^{44.62}\,{\rm erg~s^{-1}} < L_{\rm bol} < 10^{46.16}\,{\rm erg~s^{-1}}$).

\section{SED fitting for continuum luminosities} \label{sec:Lcont}

 In order to establish the $L_{\rm MIR}$-based $L_{\rm bol}$ and $M_{\rm BH}$ estimators,
we use the monochromatic continuum luminosities, $\lambda L_{\lambda}$,
at 3.4\,$\mu$m and 4.6\,$\mu$m (hereafter $L_{\rm 3.4}$ and $L_{\rm 4.6}$, respectively) in the rest-frame,
which correspond to the $\it{W1}$- and $\it{W2}$-band wavelengths of the $\it WISE$ photometric system.
However, the host galaxy contamination is non-negligible in MIR luminosity,
and the host galaxy contamination can also affect 
MIR luminosity-based properties (e.g., hot dust covering factors and MIR colors; \citealt{son22}).
Especially for low-luminosity (e.g., ${\rm L5100}< 10^{44}\,{\rm erg~s^{-1}}$) quasars,
their host galaxy contributions are significant even in the optical \citep{shen11}.

 In order to estimate the host galaxy contribution,
we use the photometric data of SDSS, 2MASS PSC, and $\it{WISE}$.
Note that $\it{W4}$ fluxes can sporadically produce unreliable results
that overemphasize IR components, and thus, are excluded in this work.
We fit the photometric data, $f(\lambda)$, with a SED model.
The SED model is a weighted sum of the spectra of
AGN ($A(\lambda)$), elliptical galaxy ($E(\lambda)$),
spiral galaxy ($S(\lambda)$), and irregular galaxy ($I(\lambda)$), which is defined as
\begin{equation}
	f(\lambda) = C_{1}A(\lambda) + C_{2}E(\lambda) + C_{3}S(\lambda) + C_{4}I(\lambda), \label{eqn:SED_model}
\end{equation}
where $C_{1}$, $C_{2}$, $C_{3}$, and $C_{4}$ are the normalization constants of each component.  
The $A(\lambda)$, $E(\lambda)$, $S(\lambda)$, and $I(\lambda)$ are the reddened spectra of 
$A_{0}(\lambda)$, $E_{0}(\lambda)$, $S_{0}(\lambda)$, and $I_{0}(\lambda)$, respectively
(see below for how the spectra were reddened).
The $A_{0}(\lambda)$, $E_{0}(\lambda)$, $S_{0}(\lambda)$, and $I_{0}(\lambda)$
are the intrinsic spectral templates of AGN, elliptical galaxy, spiral galaxy, and irregular galaxy, respectively.
The spectral templates of elliptical, spiral, and irregular galaxies are adopted from \cite{assef10}.
Note that IR emissions from star-formation are also included in these templates.

 However, the AGN spectral template is adopted from \cite{krawczyk13},
since the fiducial bolometric luminosity is measured based on the spectral template in \cite{krawczyk13}.
We note that the AGN template of \cite{krawczyk13} has a similar SED shape to those several other AGN templates (e.g., \citealt{richards06}),
but differs from the AGN template of \cite{assef10}, especially at wavelengths shorter than 0.1\,$\mu$m.
However, within the 0.1--20\,$\mu$m range used in our SED fitting, the difference is not significant.
Hence, the measured $L_{\rm 3.4}$ and $L_{\rm 4.6}$ values are similar,
and these values based on \cite{krawczyk13} are $\sim$5\,\% higher than those from \cite{assef10},
corresponding to only 0.02\,dex in the logarithmic scale.

The reddened spectra are converted from the intrinsic spectra with the $E(B-V)$ values as
\begin{equation}
	\log \left( \frac{X(\lambda)}{X_{0}(\lambda)} \right) = - \frac{k(\lambda) E(B-V)}{1.086}, \label{eqn:Extinction_model}
\end{equation}
where $k(\lambda)$ is a reddening law from \cite{fitzpatrick99}
based on Galactic extinction curve under the assumption of $\it R_V=3.1$ (e.g., \citealt{weingartner01}).
Here, we note that $X(\lambda)$ and $X_{0}(\lambda)$ denote
the four kinds of reddened spectra and their intrinsic spectra, respectively.

 Here, the different dust extinction is applied to the AGN and host galaxy components,
as often done in previous studies (e.g., \citealt{assef10}).
We also perform the SED fitting by applying the same dust extinction to the AGN and host galaxy templates.
However, the difference in the measured MIR continuum luminosity is negligible ($<1\,\%$),
which is due to (i) the MIR continuum luminosity being insensitive to the dust extinction;
(ii) we only use the unobscured quasars ($E(B-V)<0.1$). 

\begin{figure*}
	\centering
	\figurenum{3}
	\includegraphics[width=\textwidth]{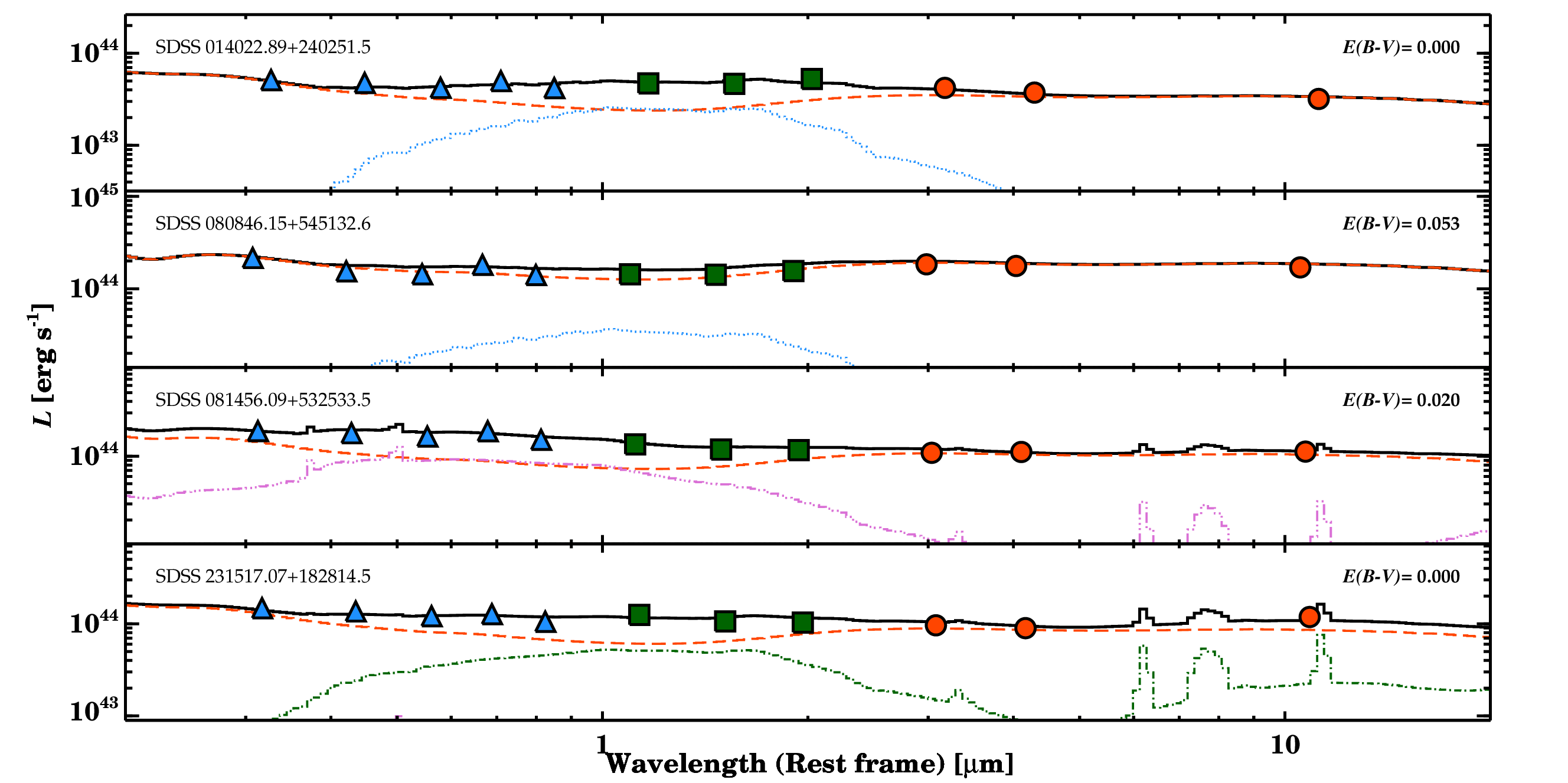}\\
	\caption{
		Photometric data of four randomly selected SDSS quasars
		with their best-fit SED models in the rest-frame.
		Blue triangles, green squares, and red circles indicate photometric data points from
		SDSS, 2MASS PSC, and $\it{WISE}$ ($\it{W1}$, $\it{W2}$, and $\it{W3}$), respectively.
		The red-dashed, blue-dotted, green-dash-dotted, and purple-dash-dot-dot-dotted lines indicate
		the reddened AGN, elliptical galaxy, spiral galaxy, and irregular galaxy spectra, respectively. 
		The black solid lines show the sums of the best-fit spectra.
		The name of the sample and the measured $E(B-V)$ value are
		presented on the top left and right side of each panel, respectively.
		\label{fig:Host_Sub}}
\end{figure*}

 This SED fitting procedure is performed
with \texttt{MPFIT} \citep{markwardt09} based on Interactive Data Language (IDL).
Figure~\ref{fig:Host_Sub} shows examples of
the photometric data with the best-fit SED model of four randomly selected SDSS quasars.

 In the following sections, we use the MIR continuum luminosities measured from the best-fit SED model.
The $L_{\rm 3.4}$ and $L_{\rm 4.6}$ values are defined as
the measured $C_{1}A_{0}(\lambda)$ values at 3.4 and 4.6\,$\mu$m in the rest-frame, respectively.

\section{Empirical relations between L5100 and $L_{\rm MIR}$} \label{sec:L5100_relation}
\begin{figure}
	\centering
	\figurenum{4}
	\includegraphics[width=\columnwidth]{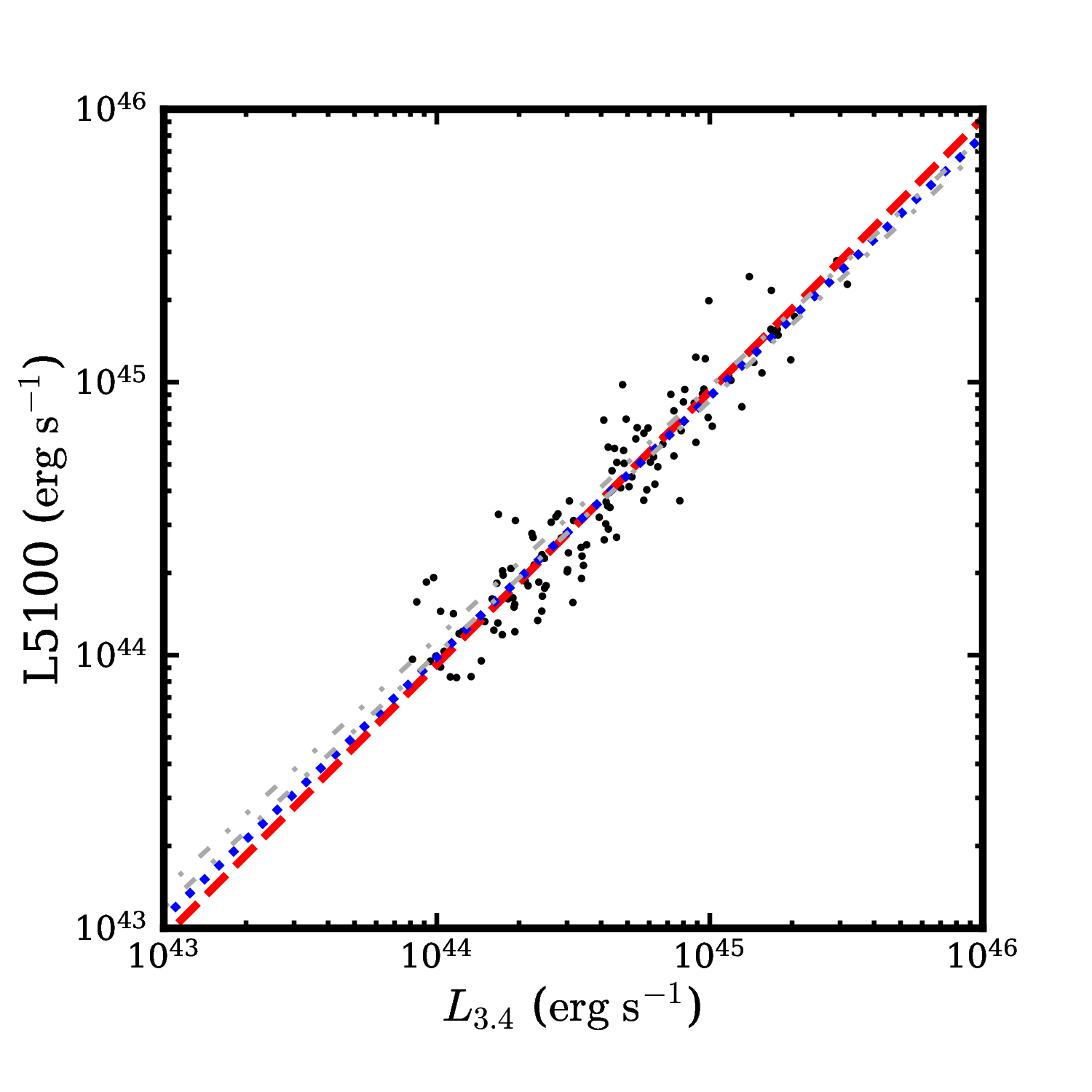}\\
	\caption{
		L5100 values versus $L_{\rm 3.4}$ values.
		The blue-dotted line represents the best-fit correlation when $\alpha$ and $\beta$ in Equation~\ref{eqn:L_relation} are set as free parameters.
		The red-dashed line denotes the best-fit correlation when the slope term $\beta$ is fixed to 1.
		Gray dashed-dotted and dashed-dot-dotted lines show the best-fit correlations derived
		from the AGN templates of \cite{richards06} and \cite{assef10}, respectively,
		where $\alpha$ and $\beta$ are treated as free parameters.
		\label{fig:L5100-LMIR}}
\end{figure}

 Before establishing $L_{\rm MIR}$-based $L_{\rm bol}$ and $M_{\rm BH}$ estimators,
we find relations between the L5100 values and the MIR continuum luminosities.
Figure~\ref{fig:L5100-LMIR} shows the correlations between the $L_{\rm 3.4}$ values and the L5100 values.
To derive the correlations between the two quantities, we assume a nonlinear relationship as
\begin{equation}
	\log \left( \frac{L_{\rm Y}}{\rm 10^{44}\,erg~s^{-1}} \right) = 
	{\rm \alpha} + {\rm \beta} \log \left( \frac{L_{\rm X}}{\rm 10^{44}\,erg~s^{-1}} \right).\label{eqn:L_relation}
\end{equation}
Here, X and Y are identifiers of the $L_{\rm MIR}$ and L5100, respectively.
We linearly fit the two quantities in the log--log plane with two free variables,
the Y intercept ($\alpha$) and the slope ($\beta$).
These correlations are fitted by using the \texttt{CURVE$\_$FIT} procedure of the \texttt{SciPy} package \citep{virtanen20}.
The fitting results are summarized in Table~\ref{tbl:L5100},
and this table also lists the root mean square (rms) scatters of the data with respect to the best-fit results.

Since the $L_{\rm 3.4}$ and the $L_{\rm 4.6}$ are derived based on the same AGN template \citep{krawczyk13},
the derived $L_{\rm 3.4}$ values are simply 1.04 times bigger than the $L_{\rm 4.6}$ value.
As a consequence, the fitting results for the $L_{\rm 4.6}$ have the same $\beta$ and rms scatter values as for the $L_{\rm 3.4}$ case.

\begin{deluxetable*}{cccccc}
	\tabletypesize{\scriptsize}
	\tablecolumns{6}
	\tablewidth{0pt}
	\tablenum{1}
	\tablecaption{Coefficients of the empirical relations between L5100 and $L_{\rm MIR}$ values\label{tbl:L5100}}
	\tablehead{
		\colhead{No.}&	\colhead{Y}&	\colhead{X}&	\colhead{$\alpha$}&	\colhead{$\beta$}& \colhead{rms}\\
		\colhead{}&		\colhead{}&		\colhead{}&		\colhead{}&			\colhead{}&		\colhead{(dex)}
	}
	\startdata
	A&	L5100&			$L_{\rm 3.4}$&	-0.034$\pm$0.012&	1 (fixed)&					0.131\\
	B&	L5100&			$L_{\rm 4.6}$&	-0.017$\pm$0.012&	1 (fixed)&					0.131\\
	C&	L5100&			$L_{\rm 3.4}$&	-0.008$\pm$0.022&	0.956$\pm$0.032&	0.130\\
	D&	L5100&			$L_{\rm 4.6}$&	0.008$\pm$0.022&	0.956$\pm$0.032&	0.130\\
	\enddata
\end{deluxetable*}

 First, we derive the correlations between the two quantities
when the slope term $\beta$ is fixed to 1, as expected in the linear relationship.
The best-fit relations are shown as the red-dashed lines in Figure~\ref{fig:L5100-LMIR},
and the derived bolometric luminosity estimators provide reasonable fits with rms scatters of 0.13\,dex.
This result is supported by Kendall's $\tau$ test.
The Kendall's $\tau$ values in these correlations are found to be $\sim$0.78.

Second, we also find the best-fit parameters of $\alpha$ and $\beta$
when we treat the slope term $\beta$ as a free parameter.
Figure~\ref{fig:L5100-LMIR} shows the derived best-fit relations as the blue-dotted lines, and the rms scatters are 0.13\,dex.
In addition, the measured Kendall's $\tau$ values are also $\sim$0.78.
These results imply that there is a negligible improvement
when the slope term $\beta$ is set as the free parameter, since the measured $\beta$ is close to 1.

 Additionally, instead of using the AGN template of \cite{krawczyk13}, 
we find the best-fit parameters with the $L_{\rm MIR}$ values that 
were derived using the AGN templates from \cite{richards06} and \cite{assef10}.
The found best-fit correlations are shown in Figure~\ref{fig:L5100-LMIR}, 
and they are not significantly different from the results based on the AGN template of \cite{krawczyk13}.

\section{Bolometric luminosity estimators} \label{sec:Lbol_est}
\begin{figure*}
	\centering
	\figurenum{5}
	\includegraphics[width=\textwidth]{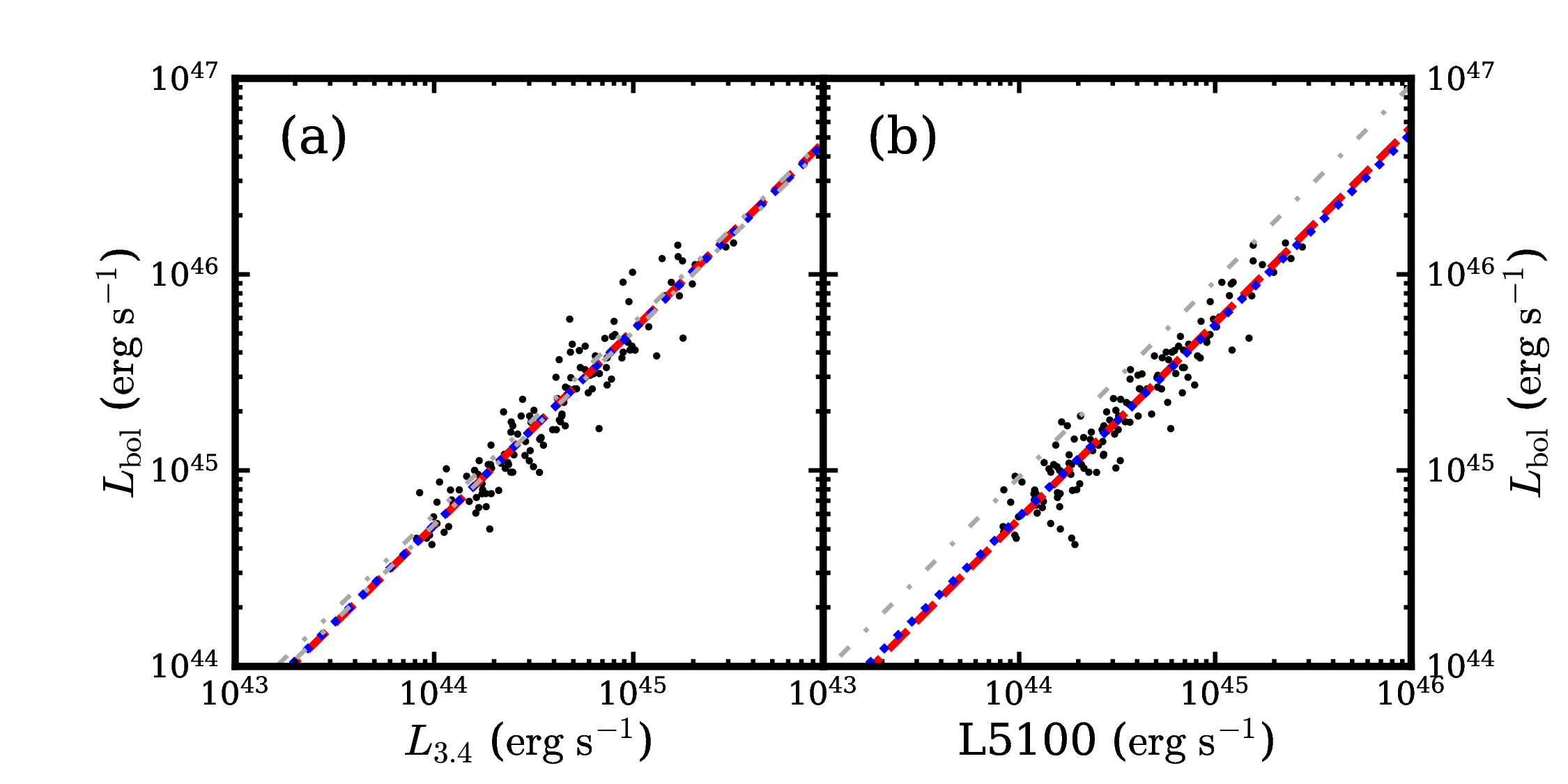}\\
	\caption{
		(a) Bolometric luminosities versus $L_{\rm 3.4}$ values.
		The bolometric luminosities are adopted from \cite{krawczyk13}.
	    The blue-dotted line represents the relation when $\alpha$ and $\beta$ in Equation \ref{eqn:L_relation} are set as free parameters,
        and the red-dashed line denotes the relation when the slope term $\beta$ is fixed to 1.
        Gray dashed-dotted and dashed-dot-dotted lines denote the best-fit correlations derived with
        the AGN templates of \cite{richards06} and \cite{assef10}, respectively,
        where $\alpha$ and $\beta$ are treated as free parameters.
    	(b) Bolometric luminosities vs. L5100 values.
    	The meanings of red-dashed and blue-dotted lines are identical to those in the left panel.
    	Gray dashed-dotted line shows the relation of L5100--$L_{\rm bol}$ by adopting the correction factor of 9.26 \citep{shen11},
    	which is measured based on the AGN template of \cite{richards06}.
    	The discrepancy between the red-dashed and blue-dotted lines is due to
    	\cite{richards06} ussing a different wavelength range (100\,$\mu$m--10\,keV) for measuring the bolometric luminosity
    	from that used (1\,$\mu$m--10\,keV) by \cite{krawczyk13}.
		\label{fig:Lbol-LMIR}}
\end{figure*}

 In this section, we find correlations between $L_{\rm bol}$ and $L_{\rm MIR}$
for deriving the $L_{\rm MIR}$-based $L_{\rm bol}$ estimators.
First, we find the $L_{\rm bol}$--$L_{\rm MIR}$ relations
when the slope term $\beta$ in Equation \ref{eqn:L_relation} is fixed to 1. 
The found best-fit relations are summarized as the parameter sets A and B in Table \ref{tbl:Lbol},
and the best-fit $L_{\rm bol}$--$L_{\rm 3.4}$ relation is shown as the red-dashed line in Figure \ref{fig:Lbol-LMIR}.
The derived $L_{\rm MIR}$-based $L_{\rm bol}$ estimators provide reasonable fits with rms scatters of 0.13\,dex,
and the Kendall's $\tau$ values are found to be $\sim$0.80.

 Second, we also find the best-fit parameters of $\alpha$ and $\beta$
when we treat the slope term $\beta$ as a free parameter.
Figure \ref{fig:Lbol-LMIR} shows the derived best-fit $L_{\rm bol}$--$L_{\rm 3.4}$ relation 
as the blue-dotted lines, and the rms scatters are 0.13\,dex. 
In addition, the measured Kendall's $\tau$ values are also $\sim$0.80. 
The best-fit relations are listed as the parameter sets C and D in Table \ref{tbl:Lbol},
and this result implies that the slope term is close to 1, even when the slope term is set as the free parameter.

Furthermore, in a manner consistent with the approach in Section~\ref{sec:L5100_relation},
we derive the best-fit $L_{\rm bol}$--$L_{\rm 3.4}$ relations by using the $L_{\rm 3.4}$ values obtained with
the AGN templates of \cite{richards06} and \cite{assef10}.
The found best-fit $L_{\rm bol}$--$L_{\rm 3.4}$ relations are not significantly different from those based on the AGN template of \cite{krawczyk13},
and they are shown in Figure~\ref{fig:Lbol-LMIR}.

 Additionally, we also find correlations between the bolometric luminosities and the L5100 values.
We find best-fit relations when the slope term $\beta$ in Equation \ref{eqn:L_relation} is fixed to 1 or treated as a free parameter,
and the derived relations are also presented as the parameter sets E and F in Table \ref{tbl:Lbol}.
Note that these results are obtained when the 1\,$\mu$m--10\,keV luminosity is adopted as the bolometric luminosity;
hence the discrepancy arises when compared to 
the bolometric luminosity correction factors (e.g., 9.26; \citealt{shen11}) 
using different wavelength ranges (e.g., 100\,$\mu$m--10\,keV; \citealt{richards06}).

 Note that the correlations of $L_{\rm bol}$--$L_{\rm MIR}$ can be affected by the hot dust covering factor.
The covering factor is known to vary from object to object \citep{roseboom13,kim15b},
and the distribution of the covering factor is expected to manifest as the scatters of the found relations in Table~\ref{tbl:Lbol}.
Moreover, there is still a debate on the relation between the covering factor and the luminosity (e.g., \citealt{mor11,roseboom13,kim15b}).
For the case that the covering factor decreases with the luminosity increasing (e.g., \citealt{mor11,roseboom13}),
we expect the slope term $\beta$ 
to be larger than the case where the dust covering factor is uniform, 
but the presence of the covering factor effects are not evident in the $L_{\rm bol}$-$L_{\rm MIR}$ relationships found.

\begin{figure*}
	\centering
	\figurenum{6}
	\includegraphics[width=\textwidth]{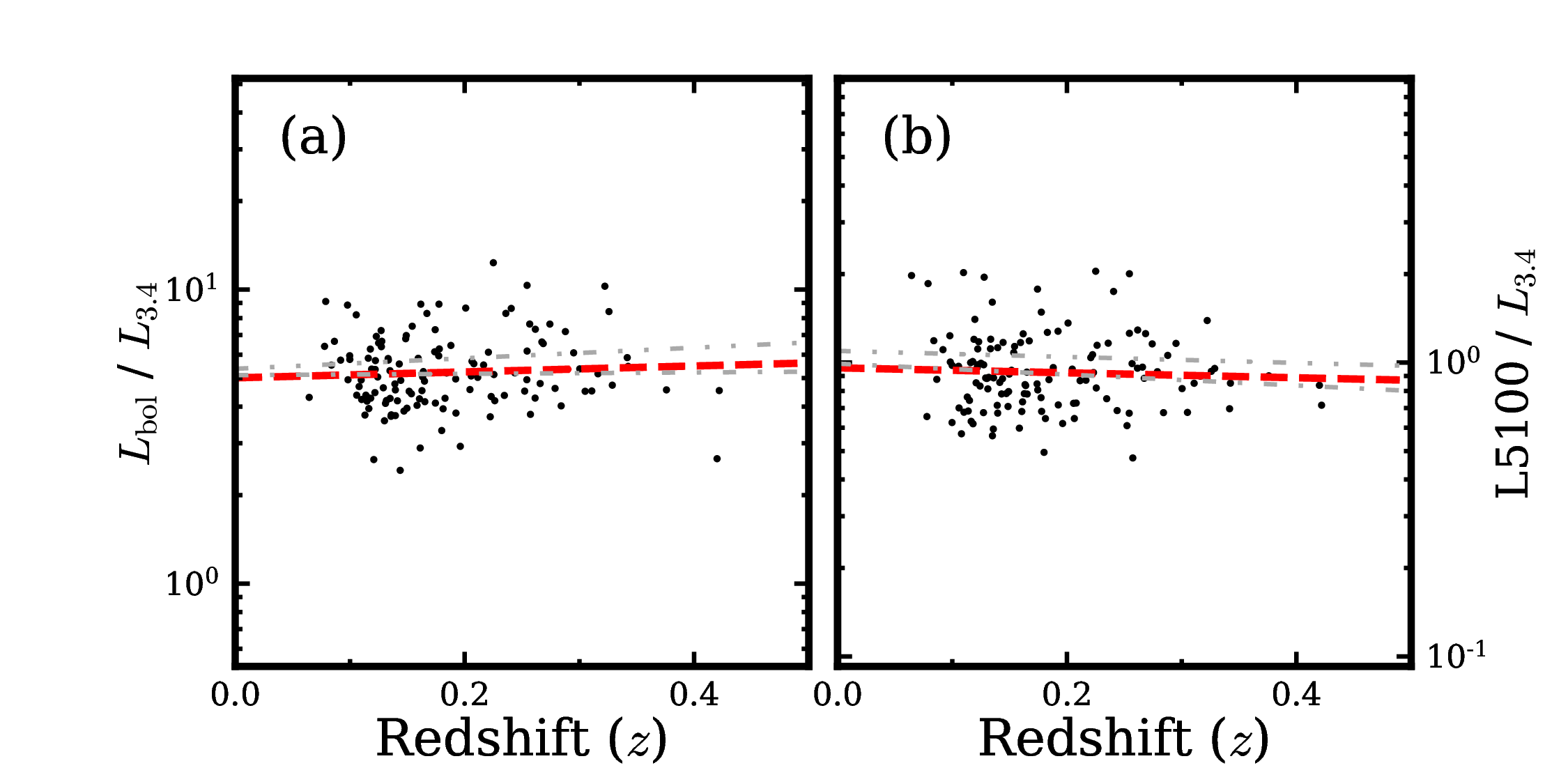}\\
	\caption{
		(a) Comparison between $L_{\rm bol}$/$L_{\rm 3.4}$ and redshift.
		The red-dashed line denotes the best-fit relationship 
		when $\alpha$ and $\beta$ in Equation~\ref{eqn:Lratio} are set as free parameters.
		The gray dashed-dotted and dashed-dot-dotted lines represent the best-fit correlations derived with
		the AGN templates of \cite{richards06} and \cite{assef10}, respectively,
		where all parameters are treated as free parameters.
		(b) L5100/$L_{\rm 3.4}$ vs. redshift.
		The meanings of the lines are identical to those in the left panel.
		\label{fig:Lratio-redshift}}
\end{figure*}

 Furthermore, we show that the relations of $L_{\rm 3.4}$--$L_{\rm bol}$ and $L_{\rm 3.4}$--L5100 
have no dependence on the redshift.
To find the redshift dependence, we assumed a relationship as
\begin{equation}
	\log \left( L/L_{\rm 3.4} \right) = \alpha+\beta\times{\rm z}, \label{eqn:Lratio}
\end{equation}
where $z$ is redshift, and $L$ is $L_{\rm bol}$ or L5100.
We fit the data by setting $\alpha$ and $\beta$ as free parameters.
In the relationship with redshift, both the $L_{\rm bol}$/$L_{\rm 3.4}$ and L5100/$L_{\rm 3.4}$ ratios have slope terms close to 0.
Moreover, these results are consistent even when using the AGN templates of \cite{richards06} and \cite{assef10}
instead of the AGN template of \cite{krawczyk13}.
These results are shown in Figure~\ref{fig:Lratio-redshift}.

\begin{deluxetable*}{cccccc}
	\tabletypesize{\scriptsize}
	\tablecolumns{6}
	\tablewidth{0pt}
	\tablenum{2}
	\tablecaption{Parameters of bolometric luminosity estimators \label{tbl:Lbol}}
	\tablehead{
		\colhead{No.}&	\colhead{Y}&	\colhead{X}&	\colhead{$\alpha$}&	\colhead{$\beta$}& \colhead{rms}\\
		\colhead{}&		\colhead{}&		\colhead{}&		\colhead{}&			\colhead{}&	\colhead{(dex)}
	}
	\startdata
	A&	$L_{\rm bol}$&	$L_{\rm 3.4}$&	0.718$\pm$0.011&	1 (fixed)&	0.126\\
	B&	$L_{\rm bol}$&	$L_{\rm 4.6}$&	0.735$\pm$0.011&	1 (fixed)&	0.126\\
	C&	$L_{\rm bol}$&	$L_{\rm 3.4}$&	0.722$\pm$0.021&	0.993$\pm$0.031& 0.126\\
	D&	$L_{\rm bol}$&	$L_{\rm 4.6}$&	0.739$\pm$0.021&	0.993$\pm$0.031& 0.126\\
	\hline
	E&	$L_{\rm bol}$&	L5100&	0.752$\pm$0.011&	1 (fixed)&	0.123\\
	F&	$L_{\rm bol}$&	L5100&	0.764$\pm$0.019&	0.978$\pm$0.029&	0.123\\
	\hline
	\enddata
\end{deluxetable*}

\section{BH mass estimators} \label{sec:MBH_est}
 We derive BH mass estimators based on
the MIR continuum luminosities with Balmer and Paschen line FWHM values
over the BH mass range of $10^{7.14}$--$10^{9.69}\,M_{\rm \odot}$.
Under the assumption that the BLRs are virialized, 
the BH mass can be measured as
\begin{equation}
	M_{\rm BH} = \it{f}\frac{R_{\rm BLR} \Delta V^{2}}{G},\label{eqn:MBH1}
\end{equation}
where $\it f$ is a scale factor depending on the geometry of the BLR,
$R_{\rm BLR}$ is the radius of the BLR,
and $\Delta V$ is the BLR velocity dispersion.
Throughout this study, we use the scale factor of 5.5 (e.g., \citealt{onken04}).
The $R_{\rm BLR}$ can be estimated directly by reverberation mapping experiment (e.g., \citealt{blandford82}),
but this experiment is difficult and expensive in time to measure.
Therefore, the BH masses have been alternatively estimated from a single-epoch spectrum
using empirical relations between $R_{\rm BLR}$ and continuum luminosity (e.g., \citealt{kaspi00}).

 We derive $L_{\rm MIR}$-based $M_{\rm BH}$ estimators that are analogous to the previous BH mass estimators
based on the single-epoch method (e.g., \citealt{greene05,vestergaard06,kim10,kim15b}).
For the $L_{\rm MIR}$-based $M_{\rm BH}$ estimators,
the MIR continuum luminosities serve as the $R_{\rm BLR}$,
and the Balmer broad-line FWHM values are used as the velocity dispersion term in Equation~\ref{eqn:MBH1}.
Mathematically, we need to find three unknown parameters, $\alpha$, $\beta$, and $\gamma$, in the following equation:
\begin{equation}
	\begin{aligned}
	\log \left( \frac{M_{\rm BH}}{M_{\odot}} \right) = \rm{\alpha} + \rm{\beta} \log \left( \frac{{\it L_{\rm X}}}{\rm 10^{44}\,erg~s^{-1}} \right) \\
	+ \rm{\gamma} \log \left( \frac{\rm FWHM_{\rm Y}}{\rm 10^{3}\,km~s^{-1}} \right).\label{eqn:MBH2}
	\end{aligned}
\end{equation}
Here, X and Y are the luminosity and line identifiers, respectively.
Under the virial theorem, $\gamma$ is fixed to 2, 
and $\beta$ is theoretically and empirically expected to be 0.5 (e.g., \citealt{dibai77,kaspi00,kim10}).
We adopt the fiducial BH masses from \cite{rakshit20}
who measured the BH masses based on optical properties using the $M_{\rm BH}$ estimator from \cite{vestergaard06}.
\cite{vestergaard06} established the BH mass estimator based on the L5100 and $\rm FWHM_{H\beta}$
with $\alpha$, $\beta$, and $\gamma$ of 6.91, 0.5, and 2, respectively.

\begin{deluxetable*}{cccccccc}
	\tabletypesize{\scriptsize}
	\tablecolumns{8}
	\tablewidth{0pt}
	\tablenum{3}
	\tablecaption{Parameters of BH mass estimators\label{tbl:MBH}}
	\tablehead{
		\colhead{No.}&	\colhead{X}&	\colhead{Y}&	\colhead{$\alpha$}&	\colhead{$\beta$}&	\colhead{$\gamma$}&	\colhead{L5100--$L_{\rm MIR}$ relation}& \colhead{rms}\\
		\colhead{}&		\colhead{}&		\colhead{}&		\colhead{}&			\colhead{}&		\colhead{}&	 \colhead{No. in Table~\ref{tbl:L5100}}&		\colhead{(dex)}
	}
	\startdata
	A&	$L_{\rm 3.4}$&	H$\beta$&	6.893$\pm$0.021&	0.5 &	2 &	A& 0.066\\
	B&	$L_{\rm 4.6}$&	H$\beta$&	6.902$\pm$0.021&	0.5 &	2 &	B& 0.066\\
	C&	$L_{\rm 3.4}$&	H$\beta$&	6.906$\pm$0.023&	0.478$\pm$0.016&	2&	C& 0.065\\
	D&	$L_{\rm 4.6}$&	H$\beta$&	6.914$\pm$0.023&	0.478$\pm$0.016&	2&	D& 0.065\\
	\hline
	E&	$L_{\rm 3.4}$&	H$\alpha$&	6.952$\pm$0.059&	0.5&							2.06$\pm$0.06& A&	0.204\\
	F&	$L_{\rm 4.6}$&	H$\alpha$&	6.960$\pm$0.059&	0.5&							2.06$\pm$0.06& B& 	0.204\\
	G&	$L_{\rm 3.4}$&	H$\alpha$& 6.964$\pm$0.060&		0.478$\pm$0.016&	2.06$\pm$0.06& C&	0.204\\
	H&	$L_{\rm 4.6}$&	H$\alpha$&	6.973$\pm$0.060&	0.478$\pm$0.016&	2.06$\pm$0.06&	D&	0.204\\
	\hline
	I &	$L_{\rm 3.4}$&	P$\beta$&		7.119$\pm$0.069&		0.5&	1.790$\pm$0.136&	A&	0.266\\
	J&	$L_{\rm 4.6}$&	P$\beta$&		7.128$\pm$0.069&		0.5&	1.790$\pm$0.136&	B& 0.266\\
	K&	$L_{\rm 3.4}$&	P$\alpha$&		7.103$\pm$0.077&		0.5&	2.034$\pm$0.160&	A&	0.478\\
	L&	$L_{\rm 4.6}$&	P$\alpha$&		7.112$\pm$0.077&		0.5&	2.034$\pm$0.160&	B&	0.478\\
	\enddata
\end{deluxetable*}

\subsection{$L_{\rm MIR}$-based $M_{\rm BH}$ estimators with Balmer line FWHM values} \label{sec:MBH_est_H}
 First, we derive $L_{\rm MIR}$-based $M_{\rm BH}$ estimators with the $\rm FWHM_{H\beta}$
by simply replacing the L5100 of the BH mass estimator of \cite{vestergaard06} with the MIR continuum luminosities.
In order to replace the L5100 of the BH mass estimator,
we use the L5100--$L_{\rm MIR}$ relations in Table~\ref{tbl:L5100}.

\begin{figure*}
	\centering
	\figurenum{7}
	\includegraphics[width=\textwidth]{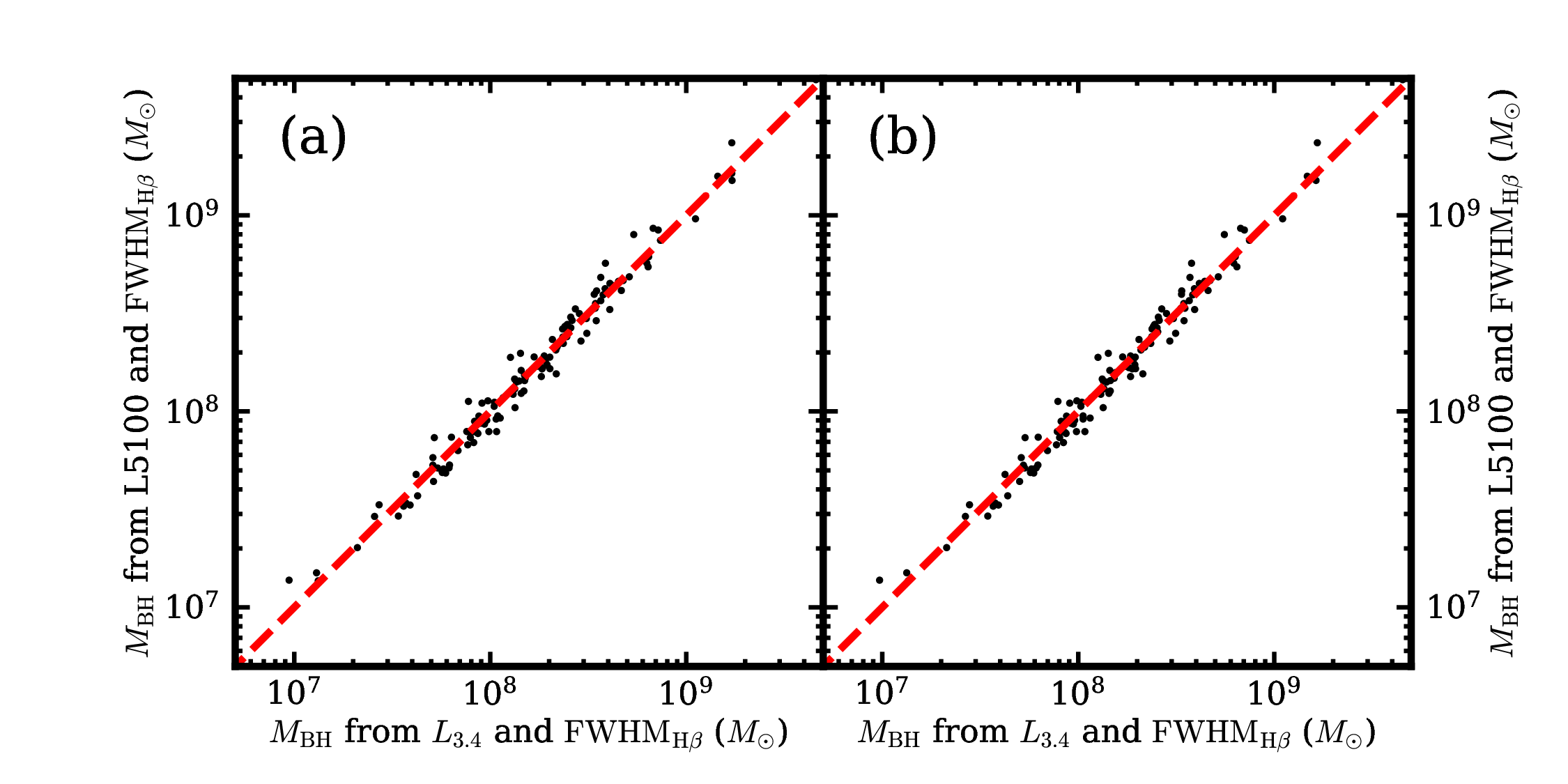}\\
	\caption{
		(a) BH masses measured from L5100 and $\rm FWHM_{H\beta}$ versus
		$M_{\rm BH}$ values from $L_{\rm 3.4}$ and $\rm FWHM_{H\beta}$.
		The $L_{\rm MIR}$-based $M_{\rm BH}$ estimator is derived by
		replacing the L5100 of the $M_{\rm BH}$ estimator of \cite{vestergaard06} with the $L_{\rm 3.4}$.
		For replacing the L5100 of the 	$M_{\rm BH}$ estimator,
		we use the linear relation of L5100--$L_{\rm MIR}$ presented as the parameter set A in Table~\ref{tbl:L5100}.
		The red-dashed line represents the case where the BH masses are identical.
		(b) Comparison of $M_{\rm BH}$ values derived from the optical properties and the $L_{\rm 3.4}$-based BH masses.
		The $L_{\rm 3.4}$-based BH masses are measured as the left panel,
		but the parameter set C in Table~\ref{tbl:L5100} is used for the L5100--$L_{\rm MIR}$ relation.
		The meaning of the red-dashed line is identical to that in the left panel.
		\label{fig:MBH-LMIR}}
\end{figure*}

 We obtain the $L_{\rm MIR}$-based BH mass estimators
when we use the parameter sets A and B in Table~\ref{tbl:L5100} as the L5100--$L_{\rm MIR}$ relation.
We find that these $L_{\rm MIR}$-based BH mass estimators reproduce the $M_{\rm BH}$
with rms scatters of $\sim$0.07\,dex.
The derived $L_{\rm MIR}$-based BH mass estimators are presented as the parameter sets A and B in Table~\ref{tbl:MBH},
and Figure~\ref{fig:MBH-LMIR} shows the comparisons of
the $M_{\rm BH}$ values based on the L5100 \citep{vestergaard06}
and the $L_{\rm 3.4}$-based BH masses.
Note that we omit to show the comparison with the $L_{\rm 4.6}$-based BH masses,
since the $L_{\rm 3.4}$- and $L_{\rm 4.6}$-based BH mass estimators
have the exact same $\beta$, $\gamma$, and scatter.

 Moreover, we obtain the $L_{\rm MIR}$-based BH mass estimators
when the parameter sets C and D in Table~\ref{tbl:L5100} is adopted as the L5100--$L_{\rm MIR}$ relation.
The rms scatters from the $L_{\rm MIR}$-based BH masses are $\sim$0.07\,dex,
and the $L_{\rm MIR}$-based BH mass estimators are presented as the parameter sets C and D in Table~\ref{tbl:MBH}.
Figure~\ref{fig:MBH-LMIR} also shows the comparisons of
the $M_{\rm BH}$ values from the optical properties \citep{vestergaard06} and the above $L_{\rm MIR}$-based BH masses.

 Second, we also obtain the $L_{\rm MIR}$-based $M_{\rm BH}$ estimators with the $\rm FWHM_{H\alpha}$
by replacing the $\rm FWHM_{H\beta}$ of the found $L_{\rm MIR}$-based $M_{\rm BH}$ estimators with the $\rm FWHM_{H\alpha}$,
and the parameter sets A, B, C, and D in Table~\ref{tbl:MBH} are used as the $\rm FWHM_{H\beta}$-based $M_{\rm BH}$ estimators.
For replacing the $\rm FWHM_{H\beta}$, we adopt the $\rm FWHM_{H\beta}$--$\rm FWHM_{H\alpha}$ relation in \cite{greene05}, which is 
\begin{equation}
	 \frac{\rm FWHM_{H\beta}}{\rm 1000\,km~s^{-1}} = 
	\rm{\left( 1.07 \pm 0.07 \right)} \times \left( \frac{\rm FWHM_{H\alpha}}{\rm 1000\,km~s^{-1}} \right)^{ \left( 1.03 \pm 0.03 \right) }.
\end{equation}
The derived $L_{\rm MIR}$-based $M_{\rm BH}$ estimators with the $\rm FWHM_{H\alpha}$
are presented as the parameter sets E, F, G, and H in Table~\ref{tbl:MBH},
and they can reproduce the BH masses with rms scatters of $\sim$0.20\,dex
that is dominated by the rms scatter in the $\rm FWHM_{H\beta}$--$\rm FWHM_{H\alpha}$ relation \citep{greene05}.

\subsection{$L_{\rm MIR}$-based $M_{\rm BH}$ estimators with Paschen line FWHM values} \label{sec:MBH_est_P}
 Here, we obtain BH mass estimators
by replacing the $\rm FWHM_{H\beta}$ of the $L_{\rm MIR}$-based $M_{\rm BH}$ estimators with the Paschen line FWHM values.
Some extremely dust-obscured quasars have only narrow line components in their Balmer lines,
but their Paschen lines can have broad-line components.
Hence, we expect that the BH mass estimators with the Paschen line FWHM values
can be applied for studying heavily obscured quasars 
when their IR spectroscopic data are obtained.

 For deriving the BH mass estimators, we need the Paschen line FWHM values.
The Paschen line FWHM values can be obtained by using empirical relations
between the Balmer and Paschen line FWHM values.
\cite{kim10} found the empirical relations with 37 unobscured type 1 AGNs at low-$z$ ($z < 0.4$)
by performing a linear bisector fit, 
and the found relations were presented in the form of
\begin{equation}
	\log \left( \frac{\rm FWHM_{H\beta}}{\rm 1000\,km~s^{-1}} \right) = 
	\rm{\alpha} + \rm{\beta} \log \left( \frac{\rm FWHM_{Paschen}}{\rm 1000\,km~s^{-1}} \right).
\end{equation}
The derived $\alpha$ and $\beta$ values for the P$\beta$ line are
0.113$\pm$0.033 and 0.895$\pm$0.068,
and the P$\alpha$ line are 0.105$\pm$0.037 and 1.017$\pm$0.080, respectively.

Using the relations between the $\rm FWHM_{H\beta}$ and $\rm FWHM_{Paschen}$,
we convert the parameter sets A and B in Table~\ref{tbl:MBH}
by replacing the $\rm FWHM_{H\beta}$ with the $\rm FWHM_{P\beta}$ and the $\rm FWHM_{P\alpha}$.
Finally, we obtain four BH mass estimators with the Paschen line FWHM values.
The obtained BH mass estimators are expressed as Equation~\ref{eqn:MBH2},
and the derived $\alpha$, $\beta$, and $\gamma$ coefficients are listed 
as the parameter sets I, J, K, and L in Table~\ref{tbl:MBH}.

\section{Discussion}

\subsection{Host galaxy contamination} \label{sec:HF}
 In this subsection, we examine the host galaxy contamination in the $L_{\rm IR}$ values.
We measure the host contamination by fitting SED model as performed in Section~\ref{sec:Lcont},
and the host contamination at a given wavelength is defined as
\begin{equation}
	\begin{aligned}
	{\rm Host~galaxy~contamination~(\lambda) = }\\
	\frac{C_{2}E_{0}(\lambda) + C_{3}S_{0}(\lambda) + C_{4}I_{0}(\lambda)}{C_{1}A_{0}(\lambda) + C_{2}E_{0}(\lambda) + C_{3}S_{0}(\lambda) + C_{4}I_{0}(\lambda)}.
	\end{aligned}
\end{equation}

\begin{figure}
	\centering
	\figurenum{8}
	\includegraphics[width=\columnwidth]{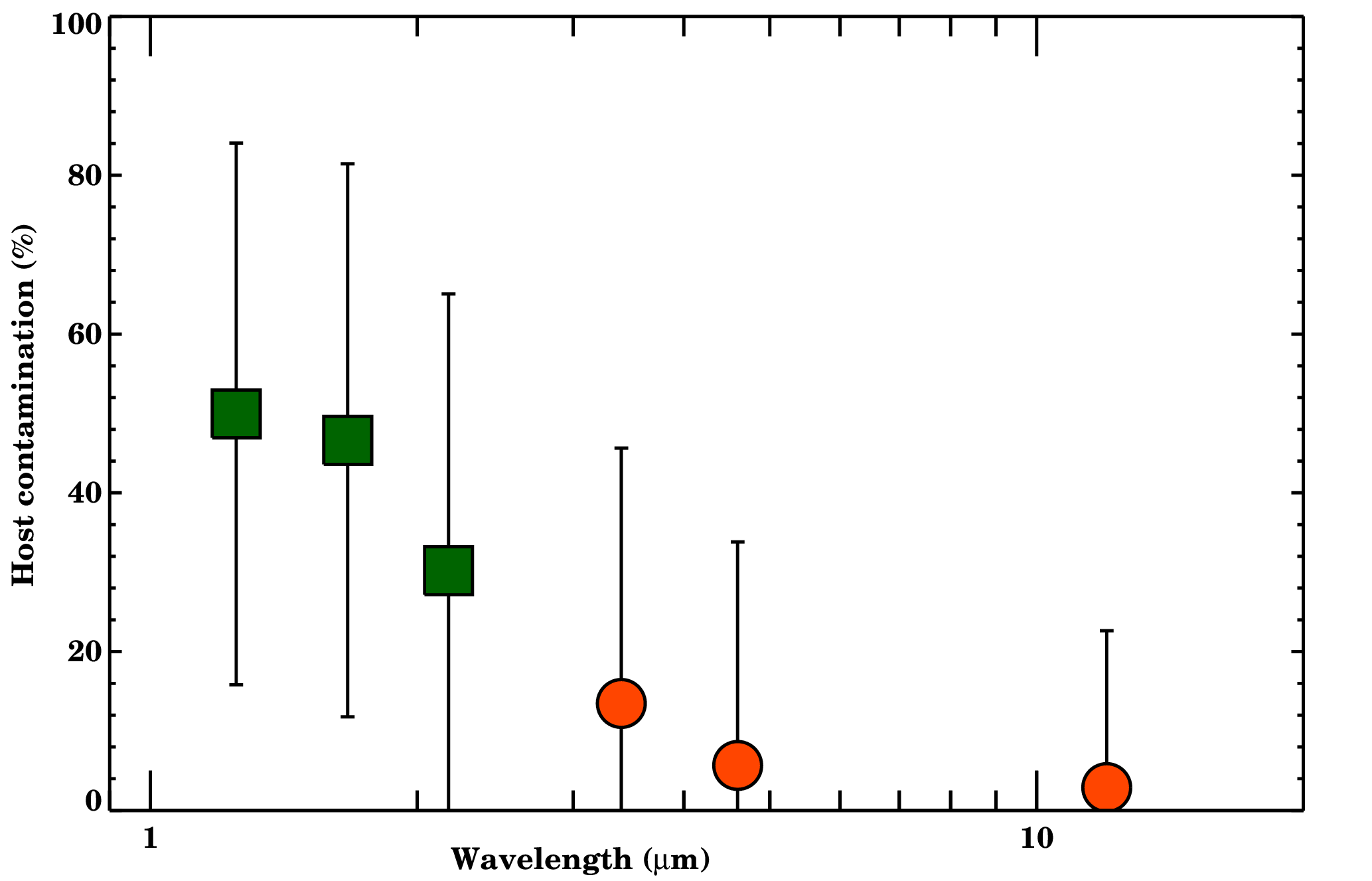}\\
	\caption{
		Host galaxy contamination at 1.2, 1.7, 2.2, 3.4, 4.6, and 12\,$\mu$m in the rest-frame.
		Green squares denote the measured host contamination at 1.2, 1.7, and 2.2\,$\mu$m,
		and red circles mean those at 3.4, 4.6, and 12\,$\mu$m.
		\label{fig:Host_frac}}
\end{figure}

 We measure the host contamination
at 1.2, 1.7, 2.2, 3.4, 4.6, and 12\,$\mu$m in the rest-frame.
Figure~\ref{fig:Host_frac} shows the median host contamination of our sample. % along with wavelength.
The host contamination is significant ($>20$\,\%; up to $\sim$50\,\%) at 1.2, 1.7, and 2.2\,$\mu$m,
but it becomes negligible ($< 20\,\%$) at 3.4, 4.6, and 12\,$\mu$m.

 The host contamination is found to be very low ($< 10$\,\%) at 12\,$\mu$m. 
Our host galaxy SED templates contain the reprocessed IR emission from star-formation, 
but the host contamination at $>12$ $\mu$m may be underestimated 
if there is a more significant star-formation activity in the host galaxy than the template galaxy SEDs. 
On the other hand, the MIR emission contribution from star-formation is relatively weak at 3.4 and 4.6\,$\mu$m, 
so we expect that the star-formation does not affect the measured host contamination at 3.4 and 4.6\,$\mu$m.

 In addition, the host contamination at 1.2, 1.7, and 2.2\,$\mu$m are found to be higher than 20\,\%.
This result remains unchanged even when using the AGN template of \cite{richards06},
which is due to that the stellar radiation peaks at these wavelengths.
Note that these measurements can increase ($\gtrsim$40\,\%; up to $\sim$70\,\%)
when the AGN template is adopted from \cite{assef10}.
The AGN SED template in \cite{assef10} is weaker at 1\,$\mu$m than other AGN templates (e.g., \citealt{elvis94,richards06,krawczyk13}),
and it is suspected to yield the overestimated host contamination at optical and NIR. 
\cite{hickox17} also reported this issue.
They measured the host contamination at 1\,$\mu$m,
and found it $\sim$75\,\% when using the AGN template of \cite{assef10}.
However, the host contamination decreases to $\sim$50--60\,\% with the AGN template of \cite{richards06}.
Moreover, \cite{richards06} measured the host contamination at 1.6\,$\mu$m, and it is found to be only $\sim$30--40\,\%.
Therefore, the host contamination measurements at optical and NIR can be somewhat increased
when using the AGN template of \cite{assef10}.

\subsection{Do SDSS quasars suffer from dust extinction?} \label{sec:SDSS_QSOs}

 In this subsection, we compare the $L_{\rm bol}$ and $M_{\rm BH}$ values
measured using the MIR continuum luminosity-based estimators to those derived based on the L5100 values
to investigate how much SDSS quasars may suffer from the dust extinction.
For this comparison, we use the sample selected from the similar criteria described in Section~\ref{sec:sample},
excepting for the Balmer decrement and $E(B-V)$ limits.
We finally take 242 low-$z$ ($z \lesssim 0.5$) SDSS quasars, and they have wide ranges of 
BH mass ($10^{7.07}\,M_{\rm \odot} < M_{\rm BH} < 10^{9.69}\,M_{\rm \odot}$)
and bolometric luminosity ($10^{44.33}\,{\rm erg~s^{-1}} < L_{\rm bol} < 10^{46.16}\,{\rm erg~s^{-1}}$).

 We measure the $L_{\rm bol}$ and $M_{\rm BH}$ values using the $L_{\rm MIR}$-based estimators
($L_{\rm bol, MIR}$ and $M_{\rm BH, MIR}$ values, hereafter).
The $L_{\rm bol, MIR}$ and $M_{\rm BH, MIR}$ values are derived by using 
the parameter set B in Table~\ref{tbl:Lbol} and B in Table~\ref{tbl:MBH}, respectively.
Furthermore, we adopt the $L_{\rm bol}$ and $M_{\rm BH}$ values based on L5100 values.
We denote them as $L_{\rm bol, opt}$ and $M_{\rm BH, opt}$ hereafter.
The $L_{\rm bol, opt}$ values were measured by using the parameter set E in Table~\ref{tbl:Lbol},
and the $M_{\rm BH, opt}$ values were measured with the L5100 and $\rm FWHM_{\rm H\beta}$ values \citep{vestergaard06}
as described in Section~\ref{sec:sample}.

\begin{figure*}
	\centering
	\figurenum{9}
	\includegraphics[width=\textwidth]{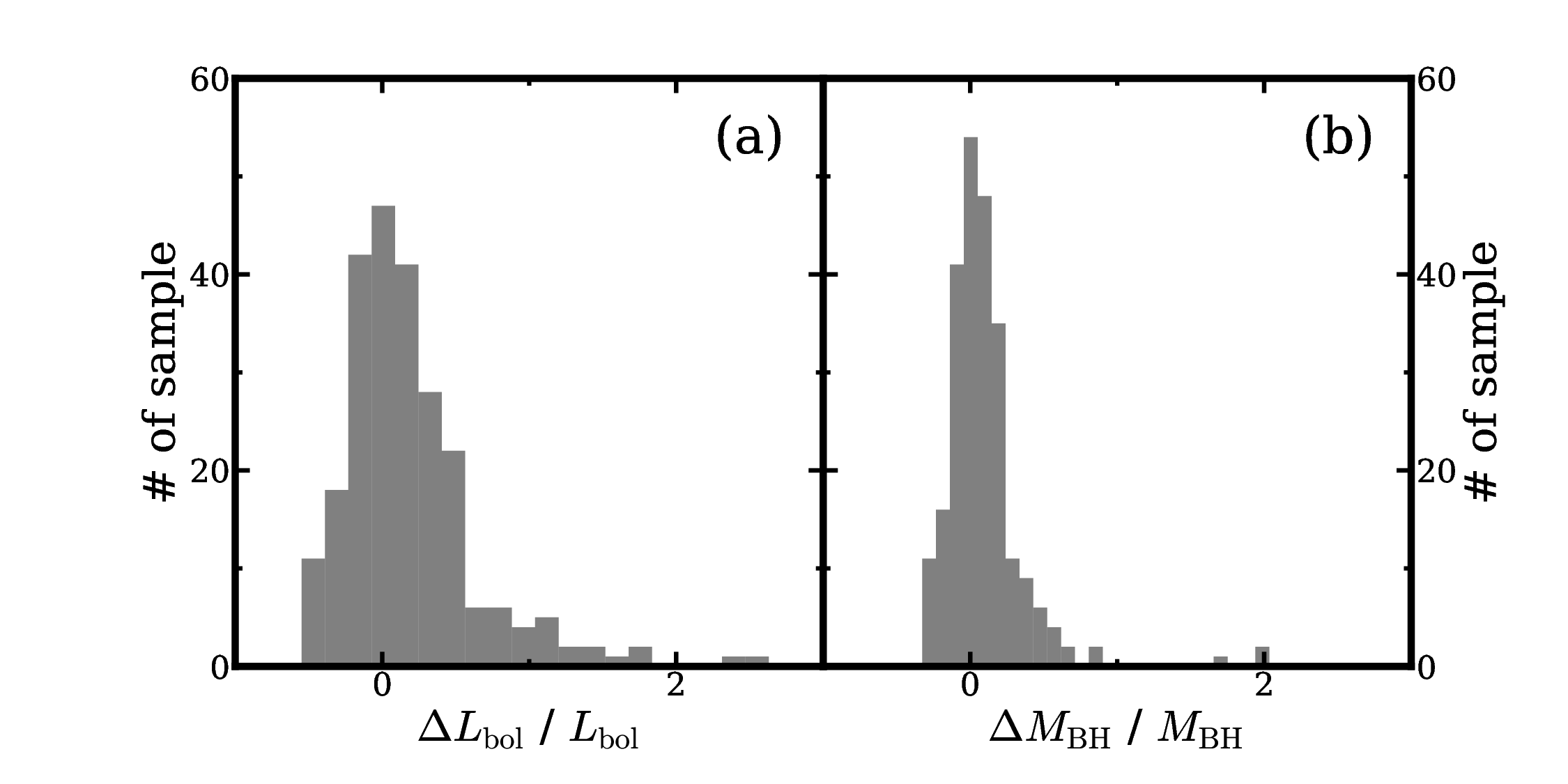}\\
	\caption{
		(a) Comparison of $L_{\rm bol}$ derived from the $L_{\rm 4.6}$ vs. 
		$L_{\rm bol}$ measured based on the L5100.
		The gray histogram represents the distribution of $\Delta L_{\rm bol} / L_{\rm bol}$, which is ($L_{\rm bol, MIR} - L_{\rm bol, opt}$)/$L_{\rm bol, opt}$,
		and its distribution is clearly skewed toward higher values.
		(b) Comparison of $M_{\rm BH}$ measured from the $L_{\rm 4.6}$ and the L5100.
		The gray histogram shows the distribution of $\Delta M_{\rm BH} / M_{\rm BH}$, which is ($M_{\rm BH, MIR} - M_{\rm BH, opt}$)/$M_{\rm BH, opt}$,
		which are somewhat skewed toward higher values.
		\label{fig:MIR-opt_dist}}
\end{figure*}

 Figure~\ref{fig:MIR-opt_dist} shows comparisons of these properties, implying that some SDSS quasars suffer from dust extinction.
A non-negligible fraction (16\,\%) of the sample possesses $L_{\rm bol, MIR} >1.5 \times L_{\rm bol, opt}$.
Considering the $L_{\rm bol, MIR}$ measurement is relatively immune from the dust extinction effects,
these $L_{\rm bol, opt}$ values are expected to be underestimated. 
For $M_{\rm BH}$, the underestimation from optical measurements is less serious
than $L_{\rm bol}$ since $M_{\rm BH}$ is proportional to $L^{0.5}$ rather than $L$ of $L_{\rm bol}$. 
Nevertheless, $M_{\rm BH, opt}$ values of 16\,\% of the SDSS quasars 
are underestimated by a factor of 1.22 ($\sim \sqrt{1.5}$) or more.  
The Eddington ratio is also proportional to $L^{0.5}$ for both optical and MIR luminosity-based estimators, 
so the optically derived Eddington ratios ($\lambda_{\rm Edd}$) are less than 80\,\% of the MIR-derived $\lambda_{\rm Edd}$
for 14\,\% of the SDSS quasars.

\begin{figure*}
	\centering
	\figurenum{10}
	\includegraphics[width=\textwidth]{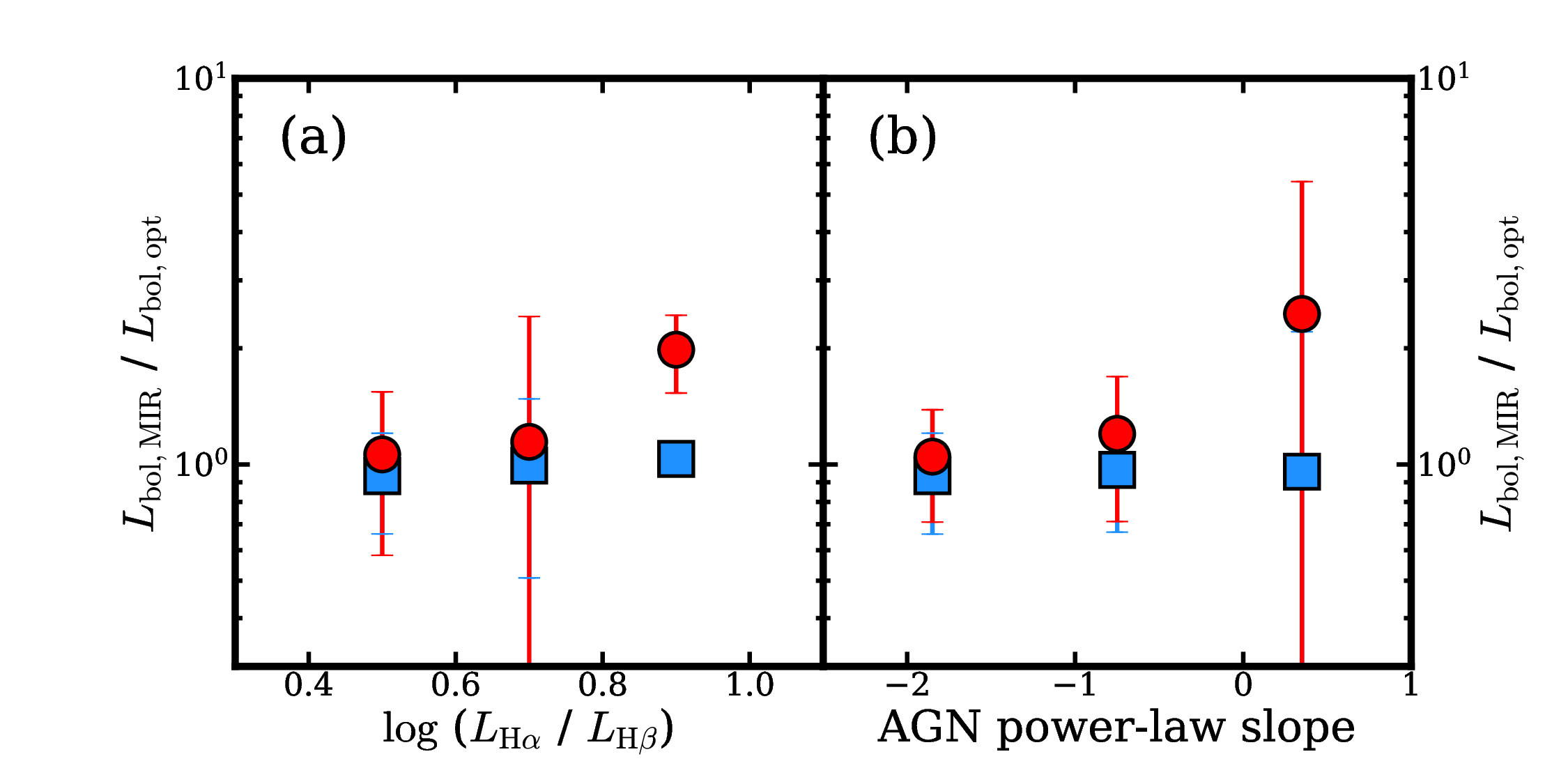}\\
	\caption{
		(a) $L_{\rm bol, MIR}$/$L_{\rm bol, opt}$ along with $\log \left( L_{\rm H\alpha} / L_{\rm H\beta} \right)$.
		Red circles represent the $L_{\rm bol, MIR}$/$L_{\rm bol, opt}$,
		and blue squares the dust extinction-corrected $L_{\rm bol, MIR}$/$L_{\rm bol, opt}$.
		(b) $L_{\rm bol, MIR}$/$L_{\rm bol, opt}$ along with AGN power-law slope.
		The meanings of red circles and blue squares are identical.
		\label{fig:MIR-opt_L}}
\end{figure*}

 Furthermore, we compare the $L_{\rm bol, MIR}$ values to the $L_{\rm bol, opt}$ values,
as functions of two dust extinction indicators,
$\log \left( L_{\rm H\alpha} / L_{\rm H\beta} \right)$ and AGN optical power-law slope,
and the results are shown in Figure~\ref{fig:MIR-opt_L}.
To clarify trends of $L_{\rm bol, MIR}$/$L_{\rm bol, opt}$,
we divide them into three bins of 
$\log \left( L_{\rm H\alpha} / L_{\rm H\beta} \right)$: 0.4$\sim$0.6, 0.6$\sim$0.8, and 0.8$\sim$1.0;
and the AGN power-law slope: -2.4$\sim$-1.3, -1.3$\sim$-0.2, and -0.2$\sim$-0.9.

 For the comparisons, the $L_{\rm H\beta}$, $L_{\rm H\alpha}$, and AGN optical power-law slope values are adopted from \cite{rakshit20}.
Here, the $L_{\rm H\beta}$ and $L_{\rm H\alpha}$ represent the H$\beta$ and H$\alpha$ broad-line luminosities, respectively.
The AGN power-law slope is measured by fitting a simple power-law function of $f_{\rm \lambda} \propto \lambda^{\alpha}$
to the optical spectra, and the fitted $\alpha$ is defined as the AGN power-law slope.

  If the dust extinction plays a significant role in the large deviation in $L_{\rm bol, MIR}$ and $M_{\rm BH, MIR}$ from the optically derived values, 
we expect that the deviation becomes stronger as the dust extinction becomes stronger as indicated by these two dust extinction indicators. 
Figure~\ref{fig:MIR-opt_L} shows the $L_{\rm bol, MIR}/L_{\rm bol, opt}$
as a function of $\log \left( L_{\rm H\alpha} / L_{\rm H\beta} \right)$ and AGN power-law slope. 
Note that the average $\log \left( L_{\rm H\alpha} / L_{\rm H\beta} \right)$ and AGN power-law slope of unobscured quasars
were found to be $\approx0.5$ (e.g., \citealt{dong08}) and $\approx$ -1.5 (e.g., \citealt{vandenberk01}), respectively.
We find that $L_{\rm bol, MIR}$/$L_{\rm bol, opt}$ increases by over two times
when $\log \left( L_{\rm H\alpha} / L_{\rm H\beta} \right) \gtrsim$ 0.8 and the AGN power-law slope is $\gtrsim$ -0.2.
Figure~\ref{fig:MIR-opt_L} clearly shows such a trend, supporting the idea that
some of the SDSS quasars suffer from the dust extinction,
and their $L_{\rm bol, opt}$ and $M_{\rm BH, opt}$ values can be underestimated without applying dust extinction correction.

 We note that low-luminosity quasars and intermediate-type AGNs also have
the high $L_{\rm H\alpha} / L_{\rm H\beta}$ and AGN power-law slope.
However, the SDSS quasars are bright ($10^{44.33}\,{\rm erg~s^{-1}} < L_{\rm bol} < 10^{46.16}\,{\rm erg~s^{-1}}$) type 1 quasars.
A significant majority ($\sim$90\,\%) of the SDSS quasars have
the broad component ($\rm FWHM > 2000\,{\rm km~s^{-1}}$) in the H$\beta$ and H$\alpha$ lines.
Therefore, the dust extinction is strongly suspected to yield the trends found in Figure~\ref{fig:MIR-opt_L}.

 We also check if dust extinction correction can bring
$L_{\rm bol, MIR}$/$L_{\rm bol, opt}$ to the values close to one.
In order to correct the dust extinction, we use the $E(B-V)$ values measured by the SED fitting as described in Section~\ref{sec:Lcont},
and the dust reddening law \citep{fitzpatrick99} under the assumption of $\it R_V=3.1$ (e.g., \citealt{weingartner01}).
After correcting the dust extinction for the $L_{\rm bol, opt}$, the newly measured $L_{\rm bol, MIR}$/$L_{\rm bol, opt}$ is found to be $\sim 1$,
even at $\log \left( L_{\rm H\alpha} / L_{\rm H\beta} \right) \gtrsim 0.8$ and AGN power-law slope $\gtrsim$ -0.2.
This result also demonstrates the importance of dust extinction correction when deriving key AGN physical quantities,
just like dust extinction needs to properly taken into account when deriving star-formation rates of dusty galaxies (e.g., \citealt{shim13}). 
To first approximation, a rough correction based on $E(B-V)$ values from SED fittings may be sufficient, 
although we expect that our MIR estimates provide more accurate values of $L_{\rm bol}$ and $M_{\rm BH}$ when $E(B-V)$ values are uncertain.

 Note that we will provide the $L_{\rm bol}$ and $M_{\rm BH}$ values
derived from the MIR continuum luminosity-based estimators
for SDSS quasars in a future work.
We expect that the newly provided $L_{\rm bol}$ and $M_{\rm BH}$ values will be useful 
for investigating the nature of quasars without the effects of dust extinction.

\section{summary}\label{summary}
 We derive the $L_{\rm bol}$ and $M_{\rm BH}$ estimators
using the MIR continuum luminosities, $L_{\rm 3.4}$ and $L_{\rm 4.6}$.
Since these MIR continuum luminosities are 
(i) relatively immune from dust extinction effects;
and (ii) measurable from IR photometric data 
that are largely available in public domain,
we expect the $L_{\rm MIR}$-based $L_{\rm bol}$ and $M_{\rm BH}$ estimators 
will be extensively used for studying various types of dust-obscured quasars.
In particular, it will be useful for interpreting the data
from current and future space IR telescopes, 
James-Webb Space Telescope (\textit{JWST}) and 
the Spectro-Photometer for the History of the Universe, Epoch of Reionization, and Ices Explorer (SPHEREx),
where the IR continua and broad lines may be detectable for obscured AGNs.

 The derived $L_{\rm MIR}$-based $L_{\rm bol}$ and $M_{\rm BH}$ estimators allow
the determinations of $L_{\rm bol}$ and $M_{\rm BH}$ at an accuracy of $\lesssim 0.2$\,dex
with respect to the values derived from optical spectra for unobscured quasars.
We also derive the $M_{\rm BH}$ estimators with the Paschen line FWHM values
for possible applications for heavily dust-obscured quasars that have no broad-line components in their Balmer lines.

 We apply the derived MIR continuum luminosity-based $L_{\rm bol}$ and $M_{\rm BH}$ estimators
for the SDSS quasars at low-$z$ ($\lesssim 0.5$).
We find that the derived $L_{\rm bol}$ values using MIR data are significantly higher
($> 1.5$ times) than
those from optical luminosity-based estimators for $\sim$15\,\% of the SDSS quasars.
Such a trend is visible for $M_{\rm BH}$ to some degree too. 
We also show that a clear correlation exists between the $L_{\rm bol}$ underestimates and the degree of dust extinction, 
suggesting the dust extinction is responsible for the $L_{\rm bol}$ underestimates.
Our results imply that a non-negligible fraction of SDSS quasars are dust obscured,
and their properties should be measured by the estimators that are relatively immune from dust extinction,
such as the MIR continuum and the IR hydrogen line \citep{kim10,kim15b,kim20} based estimators,
or at least be corrected based on the dust extinction parameters that are measurable in optical.

 There will be a wealth of MIR data available for both close, bright quasars and faint, distant quasars 
from the current and upcoming space missions such as \textit{JWST} and SPHEREx.  
Therefore, the $L_{\rm MIR}$-based estimators can be extensively applied for studying dust-obscured quasars 
with the data from such space missions.

\begin{acknowledgments}
We thank the anonymous referee for the useful comments.
This work was supported by Pusan National University Research Grant, 2020.
D.K. acknowledges the support of the National Research Foundation of Korea (NRF) grant 
(Nos. 2021R1C1C1013580 and 2022R1A4A3031306)
funded by the Korean Government (MSIT).
M.I. acknowledges the support from the National Research Foundation of Korea (NRF) grants 
Nos. 2020R1A2C3011091 and 2021M3F7A1084525, funded by the Korean Government (MSIT).
M.K. acknowledges the support by the National Research Foundation of Korea (NRF) grant (No. 2022R1A4A3031306).
Y.K. was supported by the National Research Foundation of Korea (NRF) grant funded
by the Korean Government (MSIT) (Nos. 2021R1C1C2091550 and 2022R1A4A3031306).

\end{acknowledgments}

\software{
\texttt{FM$\_$UNRED} \citep{fitzpatrick99},
\texttt{MPFIT} \citep{markwardt09},
\texttt{SciPy} \citep{virtanen20},
}

\clearpage

\end{document}